\documentclass[12pt]{article}
\usepackage{epsfig}
\makeatletter

\@addtoreset{equation}{section}
\makeatother
\newdimen\tableauside\tableauside=1.0ex
\newdimen\tableaurule\tableaurule=0.4pt
\newdimen\tableaustep
\def\phantomhrule#1{\hbox{\vbox to0pt{\hrule height\tableaurule
width#1\vss}}}
\def\phantomvrule#1{\vbox{\hbox to0pt{\vrule width\tableaurule
height#1\hss}}}
\def\sqr{\vbox{%
  \phantomhrule\tableaustep

\hbox{\phantomvrule\tableaustep\kern\tableaustep\phantomvrule\tableaustep}%
  \hbox{\vbox{\phantomhrule\tableauside}\kern-\tableaurule}}}
\def\squares#1{\hbox{\count0=#1\noindent\loop\sqr
  \advance\count0 by-1 \ifnum\count0>0\repeat}}
\def\tableau#1{\vcenter{\offinterlineskip
  \tableaustep=\tableauside\advance\tableaustep by-\tableaurule
  \kern\normallineskip\hbox
    {\kern\normallineskip\vbox
      {\gettableau#1 0 }%
     \kern\normallineskip\kern\tableaurule}%
  \kern\normallineskip\kern\tableaurule}}
\def\gettableau#1 {\ifnum#1=0\let\next=\null\else
  \squares{#1}\let\next=\gettableau\fi\next}

\def\th{\theta}

\def\cob{\delta}

\def\Tr{{\rm Tr}}

\def\ka{\kappa}
\def\hf{{1\over 2}}
\def\qu{{1\over 4}}

\def\o{\over}

\def\si{\sigma}

\def\b#1{\overline{#1}}
\def\del{\partial}
\def\wg{\wedge}
\def\lap{\Delta}
\def\bra{\langle}
\def\ket{\rangle}
\def\lf{\left}
\def\ri{\right}
\def\riya{\rightarrow}

\def\la{\lambda}

\def\h#1{\widehat{#1}}

\def\om{\omega}

\def\tens{\otimes}

\def\dag{\dagger}
\def\rt#1{\sqrt{#1}}

\def\sitarel#1#2{\mathrel{\mathop{\kern0pt #1}\limits_{#2}}}

\newcommand{\nn}{\nonumber \\}

\begin{document}

\rightline{\vbox{\hbox{hep-th/0604209}}}
\baselineskip 0.8cm
\vspace*{2.5cm}
{\LARGE Wilson Loops in ${\cal N}=4$ SYM and Fermion Droplets}
\vskip 8ex
\begin{center}
  Kazumi Okuyama and Gordon W. Semenoff\\
\vskip 2mm
  Department of Physics and Astronomy,
University of British Columbia \\
 Vancouver, BC, V6T 1Z1, Canada\\
   
\end{center}
\vskip 25mm
%
\baselineskip=6mm \centerline{Abstract}

The matrix models which are conjectured to compute the circle
Wilson loop and its correlator with chiral primary operators are
mapped onto normal matrix models. A fermion droplet picture
analogous to the well-known one for chiral primary operators is
shown to emerge in the large N limit. Several examples are
computed. We find an interesting selection rule for the correlator
of a single trace Wilson loop with a chiral primary operator. It
can be non-zero only if the chiral primary is in a representation
with a single hook.  We show that the expectation value of the
Wilson loop in a large representation labelled by a Young diagram
with a single row has a first order phase transition between a
regime where it is identical to a large column representation and
a regime where it is a large wrapping number single trace Wilson
loop.

\vskip 50mm
\noindent
April 2006
\newpage

\section{Introduction}

 The basic gauge invariant operator in Yang-Mills theory is the
Wilson loop.  It is used as a diagnostic of the dynamical behavior
of the gauge theory. The area law of the Wilson loop as a signal of
confinement is a well known example. It has also provided an
interesting tool in the AdS/CFT duality where it is the Yang-Mills
operator which is the most direct probe of fundamental
strings~\cite{Rey:1998ik,Maldacena:1998im}.

Recently, highly symmetric Wilson loops have attracted considerable
attention due the intriguing possibility that their AdS/CFT duals
are described by symmetric 3-branes and
5-branes~\cite{Drukker}-\cite{Lunin:2006xr}.  This is in analogy
with large representation chiral primary operators of ${\cal N}=4$
supersymmetric Yang-Mills theory which are interpreted as giant
gravitons~\cite{McGreevy:2000cw,Balasubramanian:2001nh,Corley:2001zk}.
In the latter case, there is a beautiful picture of ``bubbling
geometries'' which relates geometrical data of the space-time on the
string theory side of AdS/CFT to the representations of SU(N) which
are used to characterize chiral primary operators~\cite{Lin:2004nb}.

In that picture, the set of ${\small\frac{1}{2}}$-BPS chiral primary
operators, the bosonic part of whose symmetry group is $SO(4)\times
SO(4)\times R^1$, are ${\rm Tr}_R(z(x))$ where $R$ is an irreducible
representation of the GL(N,C) Lie algebra. An operator has U(1)
R-charge $J$ if the Young diagram corresponding to $R$ has $J$
boxes. Then, there is a 1-1 correspondence between Young diagrams
and Slater determinants, which are the wave-functions for quantum
states of $N$ free fermions in a harmonic potential. Due to the
Pauli exclusion principle, the fermions occupy distinct states and
in the appropriate large N limit, their semi-classical trajectories
coalesce to form an incompressible droplet in the single-fermion
phase space ~\cite{Berenstein:2004kk}. An analogy is then drawn
between the shape of this droplet and the initial data that is
necessary to specify ${\small\frac{1}{2}}$-BPS, $SO(4)\times
SO(4)\times R^1$-symmetric solutions of IIB supergravity, which are
interpreted as the background space-times in which the giant
graviton is embedded~\cite{Lin:2004nb}.

In this paper, we shall further develop the incompressible droplet
model so that it can be used to compute the circle Wilson loop,
multi-wrapped and multi-trace circle loops and correlation functions
between circle loops and chiral primary operators. We do this by
mapping the Hermitian matrix model that is conjectured to describe
the circular loop and the complex matrix model which describes the
correlation function of a circular loop and a chiral primary
operator onto  normal matrix models.  A normal matrix model is a
model for complex eigenvalues. In the large $N$ limit, the
eigenvalue density forms a droplet in the complex plane. We argue
that this droplet is identical to the one which characterizes chiral
primary operators and which describes the states of matrix quantum
mechanics~\cite{Berenstein:2004kk}. Our normal matrix model
integrals simply compute overlaps between wave-functions of the
matrix oscillator in a coherent state basis.

One of the results that we shall find is that the operator product
expansion of the fundamental representation Wilson loop for a
small circle $W_{\tableau{1}}[{\rm circle}]$ contains the chiral
primary operator ${\rm Tr}_Rz(x)$ only when $R$ is a
representation of the gauge group with a single hook, for example
$R=\tableau{4 1 1 1}$. Recall that, in the AdS/CFT correspondence,
the Wilson loop is the source of a string world-sheet and the
chiral primary operator is a graviton, or giant graviton if $R$ is
big. Our result implies that the integral of the vertex operator
that couples a graviton or giant graviton to the string
world-sheet vanishes unless the graviton has a particular
property. At present we do not understand the origin of this
selection rule, but we do observe that, as a result of it, the
Wilson loop operator acts as a probe of this aspect of the
structure of a graviton.  A single hook giant graviton is a single
spherical 3-brane extended in either the $AdS_5$ or $S_5$
direction~\cite{Corley:2001zk}.

More generally, we find that the multi-trace circle Wilson loop
operator $\left(W_{\tableau{1}}[{\rm circle}]\right)^K$ contains
giant gravitons with up to $K$ hooks. In the matrix model, this is
an exact statement, holding for all values of $N$, other parameters
and representations $R$, though the reasoning that relates it to an
exact identity for the Wilson loop is most reliable in the large $N$
limit. It is a generalization of the same known result for the
matrix elements of single trace chiral primary operators and giant
gravitons, $<{\rm Tr}z^J(x) {\rm Tr}_R(\bar z(0))>$ is non-zero only
if $R$ has a single hook~\cite{Kristjansen:2002bb}.

In the literature, there are two different descriptions of the
Wilson loop which sources a D3-brane, in ref.~\cite{Drukker} the
claim was that the large wrapping number single trace Wilson loop
whereas in ~\cite{Gomis:2006sb} the loop was one in a
representation with a long single row Young diagram. We will show
that, if $\lambda $ is large enough, and in the large $N$ limit,
these two descriptions are in fact identical. The dominant
contribution in an expansion of the character for the long row
representation in symmetric polynomials turns out to be the large
winding single trace.   We shall also observe that, if $\lambda$
is lowered beyond a certain critical value $\sim 5.5$, there is a
first order phase transition to a regime where the Wilson loop has
free energy resembling the D5-brane as described in
ref.~\cite{Yamaguchi:2006tq,Gomis:2006sb}.

The loop operator of most interest in the AdS/CFT
correspondence~\cite{Rey:1998ik,Maldacena:1998im},
\begin{equation}\label{loopdef}
W[C]~=~\frac{1}{N}{\rm Tr}{\cal P}\exp\left[\int_C d\tau
\left(iA_\mu(x(\tau))\dot
x^\mu(\tau)+\phi(x(\tau))\cdot\hat\theta|\dot
x(\tau)|\right)\right]~~,
\end{equation}
is a measure of the holonomy of a heavy W-boson which is created by
a symmetry breaking condensate of the scalar fields in ${\cal N}=4$
supersymmetric Yang-Mills theory with $\hat\theta^i=
\bra\phi^i\ket/|\bra\phi\ket|$. In IIB super-string theory, this
operator provides a source for fundamental open strings with
world-sheets bounded by the contour $C$, itself located on the
boundary of $AdS_5\times S^5$. In (\ref{loopdef}) we have written
the loop appropriate to Euclidean space which will be of most
interest to us in the following.

When $C$ is a straight line or a circle, $W[C]$ is a
${\small\small\frac{1}{2}}$-BPS operator. It is therefore
conjectured to be protected by supersymmetry. When $C$ is a straight
line, it is believed that $\bra W[{\rm straight~line}]\ket=1$. This
is supported by perturbative computations up to a few orders as well
as the strong coupling limit computed using AdS/CFT. This is also
the case for a multiply wound loop or a product of Wilson loop
operators for any array of parallel straight lines.

For the circular loop in Euclidean space, the expectation value at
large $N$ is thought to be given by the large N limit of the
Gaussian matrix integral~\cite{Erickson}
\begin{equation}
\bra W[{\rm circle}]\ket= \frac{ \int dM~\frac{1}{N}{\rm
Tr}\left(e^{M}\right)~e^{-\frac{2N}{\lambda}{\rm Tr}M^2} }
 {\int dM~ e^{-\frac{2N}{\lambda}{\rm Tr}M^2}
 }~
 \longrightarrow~\frac{2}{\sqrt{\lambda}}I_1(\sqrt{\lambda}) ~{\rm
as}~N\to\infty~,~\lambda~{\rm fixed} \label{mm}\end{equation} where
$\lambda=g_{YM}^2N$ is the 't Hooft coupling. When computed in an
expansion at weak coupling, at small $\lambda$, this ratio is equal
to the sum of ladder Feynman diagrams. It has been shown explicitly
that the first few orders in diagrams which have not been included
in the sum, those with internal loops, cancel
identically~\cite{Erickson,Plefka:2001bu,Arutyunov:2001hs}. It is
conjectured, but not proven, that such corrections cancel to all
orders and that the sum of ladder diagrams is the entire
perturbative contribution. It is possible that instantons give a
non-perturbative contribution \cite{Bianchi:2002gz,Bianchi:2001jg},
so the sum of Feynman diagrams might not give the exact amplitude.
However, in the infinite $N$, 't Hooft limit, it is plausible that
instantons are suppressed and the sum of the Feynman diagrams that
is summarized in (\ref{mm}) is indeed exact.

As further evidence for the consistency of this picture, when the
limit of large $\lambda$ is taken on the right-hand-side of
eqn.~(\ref{mm}) it agrees with the AdS/CFT computation originally
given in ref.~\cite{Maldacena:1998im}, which should be valid in that
limit. The circle can be obtained from a straight line by a
conformal transformation.  The fact that the circle and line do not
have identical expectation values has been attributed to an
anomaly~\cite{Drukker:2000rr}. In ref.~\cite{Drukker:2000rr}, they
also give a very nice argument relating, at large $\lambda$, the
${\small \frac{1}{N^2}}$ expansion of (\ref{mm}) to the moduli of
Riemann surfaces in the topological expansion in string theory.

As well as the fundamental representation loop in (\ref{mm}), the
sum of ladder diagrams for higher representation loops is given by
simply inserting the higher representation loop into the matrix
integral. A similar equation applies to multiply wound Wilson loops
or correlations of coincident loops
\begin{equation}\label{mmg}\bra\prod_i W_{w_i}[{\rm circle}]\ket=
\frac{ \int dM~\prod_i\frac{1}{N}{\rm
Tr}\left(e^{w_iM}\right)~e^{-\frac{2N}{\lambda}{\rm Tr}M^2} }
 {\int dM~ e^{-\frac{2N}{\lambda}{\rm Tr}M^2} }\end{equation}
 Here, we emphasize, that this matrix model summarizes what
 is believed to be the sum of all Feynman diagrams.  The complete set of
 ${\small}\frac{1}{2}$-BPS loops, the bosonic part of whose symmetry group is
 $SL(2,R)\times SO(3)\times SO(5)$ is
 specified by considering loops in all irreducible representations
 of the $SU(N)$ gauge group. The Schur polynomial formula can be used to
 express the expectation value of an arbitrary representation in
 terms of a sum over products of traces of multiply wound loops, the
 terms in which appear in (\ref{mmg}).

When probed from a distance much greater than its radius, $r$, a
circle Wilson loop  looks like a combination of local operators
$$W[{\rm circle}]=\bra W[{\rm circle}]\ket\cdot\sum_\Delta
\xi_\Delta\cdot(2\pi r)^\Delta{\cal O}_\Delta(0)$$ where the center
of the loop is near $x=0$. We are ignoring the slight anisotropy
which should occur in this formula.  Since ${\cal N}=4$ super
Yang-Mills theory is a conformal field theory, operators can be
classified according to their conformal dimensions, $\Delta$, along
with other quantum numbers. The series contains both primary
operators and descendants. The coefficients $\xi_\Delta$ of primary
operators can be deduced from the asymptotics of the two-point
function of the loop operator and the primary field,
\begin{equation}\label{asympt}
\frac{\bra W[{\rm circle}]{\cal O}_\Delta(x)\ket}{\bra W[{\rm
circle}]\ket}= \bra{\cal O}_\Delta(0){\cal
O}_\Delta(x)\ket\xi_\Delta\cdot(2\pi r)^\Delta+~{\rm
higher~powers~of~}1/x
\end{equation} Since the loop contains descendants as well as primary operators,
the correlator with the primary operator is given by the leading
asymptotic behavior at large $|x|$.  The descendants would appear in
higher powers of $1/|x|$.

For example, consider the chiral primary operator
$${\cal O}^{i_1i_2...i_k}(x)={\rm Tr}\left(
\phi^{i_1}(x)\phi^{i_2}(x)\ldots\phi^{i_k}(x)\right) ~~{\rm
symmetric~traceless}$$ This is a ${\small\frac{1}{2}}$-BPS operator
which has an exact conformal dimension $\Delta=k$ protected by
supersymmetry. Two- and three-point functions of chiral primary
operators are independent of the Yang-Mills coupling constant. We
are most interested in chiral primary operators of a particular
kind: if we take $\hat\theta=(1,0,...)$ in (\ref{loopdef}) and
consider
\begin{equation}\label{chirprim}{\cal O}_J(x)= {\rm
Tr}(z^J(x))~~,~~~z(x)=\phi^1(x) +i\phi^2(x)\end{equation} which has
exact conformal dimension $\Delta=J$.

The correlation function of this operator with a circle Wilson loop
located at the origin   has the form
\begin{equation}\label{loopprimcor}
\frac{\bra W[{\rm circle}] {\cal O}_J(x)\ket}{\bra W[{\rm
circle}]\ket\cdot|O_{J}|^2} = \left(\frac{2\pi
r}{4\pi^2x^2}\right)^J\cdot\xi_J ~~~{\rm as}~|x|\to\infty
\end{equation}

The normalization of the chiral primary operator is defined by its
2-point function,
\begin{equation}\bra {\cal O}_J(x){\cal O}_{J'}(0)\ket=|{\cal
O}_J|^2\delta_{JJ'}\left(\frac{1}{4\pi^2
x^2}\right)^{J}\end{equation} and can be computed using the complex
matrix integral~\cite{Kristjansen:2002bb}
\begin{equation}\label{chirprimnorm}
|{\cal O}_J|^2 \delta_{JJ'}= \frac{ \int d^2z
e^{-\frac{2N}{\lambda}{\rm Tr}\bar z z}{\rm Tr}\bar z^J{\rm
Tr}z^{J'} }{ \int d^2z e^{-\frac{2N}{\lambda}{\rm Tr}\bar z z}
}\end{equation} The constants $\xi_J$ in (\ref{loopprimcor}) were
computed in the planar, small $\lambda$ limit using perturbation
theory and in large $\lambda$ limit using the AdS/CFT correspondence
in ref.~\cite{Berenstein:1998ij}. The sum of planar ladder diagrams
contributing to (\ref{loopprimcor}) was found in
ref.~\cite{Semenoff:2001xp}:
\begin{equation}\label{ladsum}
\xi_J=\frac{1}{N}\frac{\sqrt{J\lambda}}{2^{1+\small\frac{J}{2}}}
\frac{I_J(\sqrt{\lambda})}{I_1(\sqrt{\lambda})}
\end{equation}
This was conjectured to be an exact result. The sum of planar
ladders which is summarized in (\ref{ladsum}), when extrapolated to
large $\lambda$, agrees with the AdS/CFT computation of the same
coefficients $\xi_J$ in ref.~\cite{Berenstein:1998ij}. Leading
corrections from other diagrams which are not ladders were also
computed and shown to vanish in ref.~\cite{Semenoff:2001xp},
supporting the conjecture that they give the exact result. This has
further been supported by comparison with large quantum number
limits of strings in ref.~\cite{Pestun:2002mr}.

The sum of planar ladder diagrams which contributes to
(\ref{ladsum}) is also given by the large $N$ limit of a matrix
integral. As well as the single trace (\ref{chirprim}) we could
consider a more general set of chiral primaries ${\rm Tr}_Rz(x)$
where the trace is taken in an irreducible representation.\footnote{
For an irreducible representation $R$ corresponding to a given Young
Diagram with $J$ boxes, we begin with
$$z^{i_1}_{i_2} z^{i_2}_{i_3}z^{i_3}_{i_4}...z^{i_{J-1}}_{i_J}z^{i_J}_{i_1}$$
and consider linear combinations of permutations of the upper
indices until they transform in the representation of the symmetric
group that is associated with the Young Diagram. We then contract
the indices to get a linear combination of single and multi-trace
operators. For example, when $J=2$, there are two representations $$
z^i_jz^j_i\to \frac{1}{2}\left(z^i_jz^j_i+z^j_jz^i_i\right)\to
\frac{1}{2}{\rm Tr} z^2+\frac{1}{2}{\rm Tr}z{\rm Tr} z={\rm
Tr}_{\tableau{2}}z
$$
$$z^i_jz^j_i\to \frac{1}{2}\left(z^i_jz^j_i-z^j_jz^i_i\right)\to
\frac{1}{2}{\rm Tr }z^2-\frac{1}{2}{\rm Tr}z{\rm Tr}
z={\Tr}_{\tableau{1 1}}z
$$}
In general, we could consider a Wilson loop in a representation $R$.
Its expansion in local operators could contain a chiral primary in a
representation $R'$.  The coefficient $\xi_{RR'}$ in this expansion
can also be expressed as a matrix integral,
\begin{equation}\label{mmcp}\xi_{RR'}~=~
\frac{ \int d^2z~ \frac{1}{{\rm dim}R}{\rm
Tr}_R\left(e^{{\small\frac{1}{2}} (z+\bar z)}\right)~ {\rm
Tr}_{R'}(z)~e^{-\frac{2N}{\lambda}{\rm Tr}\bar z z} } { \int d^2z~
\frac{1}{{\rm dim}R}{\rm Tr}_R \left(e^{{\small\frac{1}{2}}(z+\bar
z)}\right)~ e^{ -\frac{2N}{\lambda}{\rm Tr}\bar z z } \cdot |{\cal
O}_{R'}|^2}
\end{equation}

On the string theory side, chiral primary operators correspond to
supergravitons and their coupling to loop operators can be extracted
from asymptotics of the connected correlator of two far separated
loop operators, which are sources for fundamental string
world-sheets sitting at the boundary of $AdS_5\times S^5$. For small
representations, they should give information about the coupling of
gravitons to the fundamental string world-sheets that are sourced by
the loop operators. For bigger representations they would yield
information about the gravitational interactions of the more exotic
objects, D3-branes and D5-branes, that are sourced by the loops.
When $R$ and $R'$ are both large, (\ref{mmcp}) gives information
about the coupling of giant strings to giant gravitons.

In this Paper, we will take a closer look at some of the properties
of the matrix integrals (\ref{mm}) and (\ref{mmcp}).  In order to
study them at large $N$, we will rewrite them as models for complex
eigenvalues, which are called Normal matrix models.

\subsection{Eigenvalue models}

 In some cases, following the technique of Ginibre
\cite{ginibre}, an integral over complex matrices can be written as
an integral over complex eigenvalues where the matrices are diagonal
and, with the suitable measure, we are left with integrating over
the diagonal elements only. The most straightforward case is when
the objects inside traces in the integrand are holomorphic in the
complex matrix. An example is the integral which has to be done to
find the normalization of the chiral primary operators
\begin{equation}\label{111}
\bra{\cal O}_R{\cal O}_{R'}\ket=\frac{ \int d^2z e^{ -{ \small
\frac{2N}{\lambda} } {\rm Tr} \bar z z} {\rm Tr}_R\bar z {\rm
Tr}_{R'}z } { \int d^2z e^{ -{\small \frac{2N}{\lambda} } {\rm Tr}
\bar z z } }\end{equation} The matrices can be made upper-triangular
by a unitary transformation.   The upper-triangular property
($z_{ij}=0$ if $i>j$) is preserved by products of $z$'s and the
trace of a product of the $z$'s depends only on diagonal matrix
elements, Tr$(z^J)=\sum_i z^J_{ii}$. In the following, we will use
one index to denote diagonal components, $z_i\equiv z_{ii}$. The
conjugate matrix, $\bar z$, is lower-triangular and similar
arguments apply.

The matrix integral can thus be written as an integral over diagonal
components times the appropriate Jacobian times an integral over the
components in the upper triangle and an integration over unitary
matrices. The integration of the latter two decouples and is common
between the numerator and denominator in (\ref{111}). They cancel,
leaving the integral over diagonal components with the Jacobian. The
resulting measure is
\begin{equation}\label{measure}d^2z = \prod_{i=1}^N d^2z_i~
\Delta(z)\Delta(\bar z)\end{equation} with the vandermonde
determinant
$$
\Delta(z)=\prod_{i<j}\left( z_i-z_j\right)=\det z_i^{j-1}=\det
z_i^{N-j}
$$
When the matrix is triangular, the trace of a matrix in an
irreducible representation $R$, ${\rm Tr}_R z$ is a function only of
its diagonal components and is given by the Schur polynomial
\begin{equation}
{\rm Tr}_Rz=
{\det(z_i^{h_j+N-j})\o\det(z_i^{N-j})}={\det(z_i^{h_j+N-j})
\o\lap(z)}
\end{equation}
Here $h_1\geq h_2\cdots\geq h_N$ are the lengths of the rows of the
Young diagram associated with the representation $R$.

It is now easy to see that that (\ref{111}) can be written as an
integral for diagonal elements of matrices, and then performed
explicitly,

\begin{eqnarray}
\bra{\cal O}_R{\cal O}_{R'}\ket&=&\frac{
\int\prod_{i=1}^Nd^2z_ie^{-{\small\frac{2N}{\lambda}}|z_i|^2}~|\lap(z)|^2~
{\det(\b{z}_i^{h_j+N-j})
\o\lap(\b{z})}~{\det(z_i^{h_j'+N-j})\o\lap(z)}}
{\int\prod_{i=1}^Nd^2z_ie^{-{\small\frac{2N}{\lambda}}|z_i|^2}~|\lap(z)|^2}
\nn &=&\frac{~\sum_{\sigma,\tau\in S_N} (-1)^{{\rm deg}\sigma+{\rm
deg} \tau}~
\int\prod_{i=1}^Nd^2z_ie^{-{\small\frac{2N}{\lambda}}|z_i|^2}~ (\bar
z_i^{h_\sigma(i)+N-\sigma(i)}) (z_i^{h_{\tau(i)}'+N-\tau(i)}) }
{\sum_{\sigma,\tau\in S_N} (-1)^{{\rm deg}\sigma+{\rm deg}\tau}
~\int\prod_{i=1}^Nd^2z_ie^{-{\small\frac{2N}{\lambda}}|z_i|^2}~
(\bar z_i^{N-\sigma(i)}) ( z_i^{N-\tau(i)}) } \nn &=&\delta_{RR'}~
\prod_{j=1}^N\left(\frac{\lambda}{2N}\right)^{h_j}{(h_j+N-j)!\o(N-j)!}
\label{112}\end{eqnarray} where $S_N$ is the set of all permutations
of the integers $(1,...,N)$ and we have decomposed the determinants
into sums over permutations using the formula
$$\det M=\sum_{\sigma\in S_N}(-1)^{{\rm deg}\sigma}
\prod_{i=1}^N M_i^{\sigma(i)}$$

In the large $N$ limit, if we assume that $h_i/N\to 0$ for all
$h_i$, the result of (\ref{112}) reduces to
$$
|O_R|^2 \to \left( \frac{\lambda}{2}\right)^{\sum_i h_i}~~{\rm
as}~~N\to\infty
$$

\vskip 2cm

 In the remainder of this paper, we are going to study matrix
integrals such as (\ref{mmcp}) where making the matrices
upper-triangular does not decouple the eigenvalues. What we will
find is that, in many cases, we can nevertheless map it onto an
eigenvalue integral over complex eigenvalues, that is, a Normal
Matrix model.

\section{Normal Matrix Model}

A normal matrix model~\cite{Chau:1991gj,Chau:1997pr} is a complex
matrix model where the real and imaginary parts of the matrices are
constrained to commute with each other.  They can be simultaneously
diagonalized by conjugation with a unitary matrix, $z=U\cdot{\rm
diag}(z_1,...,z_N)U^\dagger$ and $\bar z =U{\rm diag}(\bar
z_1,...,\bar z_N)U^\dagger$ where $UU^\dagger={\cal I}$. Then, the
integration of any function of the matrices which is invariant under
$(z.\bar z)\to(U z U^\dagger, U\bar zU^\dagger)$ is taken as an
integral over the diagonal elements $z_i$ with measure given in
(\ref{measure}) and the matrix $z_{ij}$ replaced by the diagonal one
${\rm diag}(z_1,...z_N)$. Normal matrix models have recently found a
number of interesting
applications~\cite{Alexandrov:2003qk}-\cite{Teodorescu:2004ya}. They
were discussed in the context of non-critical string theory in
refs.~\cite{Ambjorn:1992xu}-\cite{Kostov:1997bn}.

The normal matrix model and the complex matrix model are equivalent
if, as is the case in (\ref{111}) solved above, aside from the
Gaussian term, the quantities in the integrand are holomorphic in
the matrices. However, in the case of the Wilson loop (\ref{mm}) or
the correlator (\ref{mmcp}), the integrand is not holomorphic and it
is less obvious that it can be written as an eigenvalue model.

Nevertheless, there are circumstances where the normal matrix model
is equivalent to the complex matrix model.  In the remainder of this
Paper, we will examine some of these examples.  The interesting
feature of a normal matrix model is that, in the large $N$ limit,
the distribution of eigenvalues has support on a two-dimensional
subset of the complex plane.  This will lead to a droplet model for
the eigenvalues.

\subsection{Fundamental representation Wilson loop}

As an example, consider the expectation value of the Wilson loop in
the fundamental representation which is given by the Hermitian
matrix integral (\ref{mm}) of the Wilson loop in this
representation,
 \begin{equation}\label{113}
 \bra W_{\tableau{1}}\ket =\frac{ \int [dM]~
 e^{-{\small\frac{2N}{\lambda}}{\rm Tr}M^2}~\frac{1}{N}{\rm
 Tr}\left(e^M\right) }{ \int [dM]~
 e^{-{\small\frac{2N}{\lambda}}{\rm Tr}M^2} }
 \end{equation}
We can easily write this as a complex matrix model.  The approach
(which we shall not follow) would be to make this into a
 complex matrix model by introducing the imaginary part as
 another Gaussian matrix $M_I$, so that, $z=M+iM_I$
 and writing
\begin{equation}\label{114}
 \bra W_{\tableau{1}}\ket =\frac{ \int [d^2z]
 e^{-{\small\frac{2N}{\lambda}}{\rm Tr}\bar z z}\frac{1}{N}{\rm
 Tr}\left(e^{{\small\frac{1}{2}}(z+\bar z)}\right) }{ \int [d^2z]
 e^{-{\small\frac{2N}{\lambda}}{\rm Tr}\bar z z} }
 \end{equation}
It is not obvious how to now write (\ref{114}) as an integral over
eigenvalues. Backtracking, we recall that we could also have written
(\ref{113}) as an integral over the eigenvalues of Hermitian
matrices,
 \begin{equation}\label{hermm1}
 \bra W_{\tableau{1}}\ket =\frac{ \int \prod_i dm_i
 e^{-{\small\frac{2N}{\lambda}}m_i^2} \Delta^2(m)~\left(e^{m_1}\right) }{  \int \prod_i dm_i
 e^{-{\small\frac{2N}{\lambda}}m_i^2} \Delta^2(m) }
 \end{equation}
 To proceed, we realize that the integral that we must do in
 (\ref{hermm1}) is formally the expectation value of the operator
 $e^{m_1}$ in the quantum state given by the wave-function $\psi(m_i)
 =  \Delta(m)e^{-{\small\frac{N}{\lambda}}\sum_i
 m_i^2}$.  This is just the ground state of a system of $N$ fermions
 moving in a Harmonic potential well. (To see this, we must recall
 some properties of determinants of polynomials.  This point is discussed in the Appendix at the end
 of this Section.  See eq.~(\ref{psisea}).).  There are
 other ways to present this integral.  One of them is uses a
 holomorphic polarization of the oscillator phase space.
 In that presentation, the ground state wave-function of an N-fermion system is $\psi(z)=\Delta(z)
 e^{-{\small\frac{N}{\lambda}}\sum_i\bar z_iz_i}$ where $z$ is a complex variable and the expectation value is

\begin{equation}\label{1160}
 \bra W_{\tableau{1}}\ket = e^{-\lambda/8N} \frac{ \int \prod_i d^2z_i
 e^{-{\small\frac{2N}{\lambda}}\bar z_i z_i} |\Delta(z)|^2~{\small\frac{1}{N}}{\rm Tr}\left(e^{{\small\sqrt{\frac{1}{2}}}
  (z+\bar z)}\right) }
 {  \int \prod_i d^2z_i
 e^{-{\small\frac{2N}{\lambda}}\bar z_i z_i} |\Delta(z)|^2 }
 \end{equation}
A detailed derivation of a general integral formula of this kind,
eq.~(\ref{theformula}) is given in the Appendix to this Section.

 This is now a normal matrix model, written as an integral over
 complex eigenvalues. Note two differences between the normal model (\ref{1160}) and
 the complex model (\ref{114}): the factor $e^{-\lambda/8N}$ in
 front of (\ref{1160}) and the factor of ${\small\frac{1}{\sqrt{2}}}$, rather
 than ${\small\frac{1}{2}}$ in front of $(z+\bar z)$.

In summary, we have found three identical presentations of the same
matrix model:
\begin{eqnarray}
\bra W_{\tableau{1}}\ket&=& {1\o Z_H}\int [dM]e^{-{2N\o\la}\Tr
M^2}{\small{1\o N}}\Tr \,e^M\quad ({\rm hermitian~model})\nn &=&{1\o
Z_C}\int [d^2z]~e^{-{2N\o\la}\Tr z\b{z}}{\small{1\o N}}\Tr
\,e^{\hf(z+\b{z})} \quad({\rm complex~model})\nn &=&{1\o
Z_N}e^{-{\la\o8N}}\int_{[z,\b{z}]=0} [d^2z]~e^{-{2N\o\la}\Tr
z\b{z}}{\small\frac{1}{N}}\Tr\, e^{{1\o\rt{2}}(z+\b{z})}\quad({\rm
normal~model}) \label{summat}
\end{eqnarray}
where we have abbreviated the denominators as the Gaussian partition
functions $Z_H,Z_C,Z_N$ for Hermitian, complex and normal matrix
models, respectively. We can easily check that the order $\la$ term
agrees in all three computations in (\ref{summat}):
\begin{eqnarray}
&&\bra W_{\tableau{1}}\ket\nn &=&1+{1\o2N}\bra \Tr\,M^2\ket_{\rm
hermitian}+\cdots=1+{1\o2N}{\la\o2N}{N^2\o2}+\cdots=1+{\la\o8}+\cdots,\nn
&=&1+{1\o4N}\bra\Tr Z\b{Z}\ket_{\rm complex}+\cdots=1+{1\o4N}{\la\o2N}{2N^2\o2}+\cdots=1+{\la\o8}+\cdots,\\
&=&1+{1\o2N}\bra\Tr Z\b{Z}\ket_{\rm normal}-{\la\o8N}+\cdots
=1+{1\o2N}{\la\o2N}{N^2+N\o2}-{\la\o8N}+\cdots =1+{\la\o8}+\cdots.
\nonumber
\end{eqnarray}
We also emphasize that the three expressions in (\ref{summat}) are
exactly equal for all values of $\la$ and $N$.

\subsection{Fermion droplet model}\label{gaussianmm}

Let us now consider the multiply wound circular Wilson loop with $k$
windings.  The discussion of the previous Section leads to the
Normal matrix integral
$$
\bra W_{\tableau{1}}^{(k)}\ket=\frac{1}{Z_N}~e^{-{\small\frac{k^2\lambda}{8N}}}~\int_{[z,\bar
z]=0} [d^2z] e^{-{\small\frac{2N}{\lambda}}{\rm Tr}\bar z z}
{\small\frac{1}{N}}{\rm Tr}e^{ {\small\frac{k}{\sqrt{2}}}(z+\bar
z)}$$ This can be written as an integral over diagonal matrices with
$$
\bra W_{\tableau{1}}^{(k)}\ket=\frac{1}{Z_N}~e^{-{\small\frac{k^2\lambda}{8N}}}~\int \prod_i
d^2z_i \exp\left(-{\small\frac{2N}{\lambda}}\sum_{i=1}^N\bar z_i
z_i+\sum_{i\neq j}\ln|z_i-z_j|^2+{\small\frac{k}{\sqrt{2}}}(z_1+\bar
z_1)\right)
$$
In the large $N$ limit, the integration over the eigenvalues
$z_2,...,z_N$ is dominated by a saddle point where the eigenvalues
satisfy the saddle point equation
$$
\frac{2}{\lambda}z_i = \frac{1}{N}\sum_{j\neq i} \frac{1}{\bar z_i -
\bar z_j}+\frac{k}{\sqrt{2}N}\delta_{i1}
$$
If we introduce the normalized eigenvalue density
$$
\rho(z) = \frac{1}{N-1}\sum_{i=2}^N \delta^2(z-z_i) ~~~,~~~ \int
d^2z \rho(z)=1
$$
the saddle point equation
\begin{equation}\label{saddpt}
\frac{2}{\lambda}z ~=~\frac{N-1}{N} \int d^2z'\frac{\rho(z')}{\bar
z-\bar z'}+\frac{1}{N}\frac{1}{\bar z-\bar z_1}~~~z\in{\rm
support~of~}\rho
\end{equation}

There are two different regimes that we can analyze:
\begin{enumerate}
\item{}{\it Small winding loop}:  $\lim_{N\to\infty}{\small\frac{k}{N}}=0$

In this regime, the last term in the saddle point equation
(\ref{saddpt}) is irrelevant and the equation is solved by the
constant density droplet
\begin{equation}\label{gaussiandensity}
\rho(z)=\left\{ \matrix{ \frac{2}{\pi\lambda} &
|z|<\sqrt{\frac{\lambda}{2}} \cr 0 & |z|>\sqrt{\frac{\lambda}{2}}
\cr } \right.
\end{equation}
There are two ways to compute the expectation value of the multiply
wound Wilson loop.  We can first simply use the formula for
evaluating the expectation value of the trace of matrices once the
density is known:
\begin{eqnarray}
\bra W_{\tableau{1}}^{(k)}\ket &=& \int d^2z  \rho(z) e^{\frac{1}{\rt{2}}k(z+\bar z) }
=\frac{2}{\pi\lambda}
\int_0^{\sqrt{\frac{\lambda}{2}}}rdr\int_0^{2\pi}d\theta
~e^{\rt{2}kr\cos\theta} \nn &=&{1\o\pi}\int_0^1 tdt\int_0^{2\pi}d\th
e^{k\rt{\la}t\cos\th} =\int_0^1 2tdt \,I_0(k\rt{\la}t)
=\frac{2}{\sqrt{\lambda}k}I_1(k\sqrt{\lambda}) \label{case1}
\end{eqnarray}
This is valid when $k<<N$ and it agrees with the Hermitian Gaussian
matrix model (\ref{mm}). \footnote{ The expectation value of the
single trace Wilson loop is known explicitly at finite $N$
\cite{Drukker}. A useful formula is a contour integral
representation \cite{Kazakov:1990nd}
\begin{equation}
\bra W_{\tableau{1}}^{(k)}\ket={2\o k\rt{\la}}\oint{d\xi\o2\pi
i}e^{{k\rt{\la}\o2}\xi} \lf({1+{k\rt{\la}\o4N\xi}\o
1-{k\rt{\la}\o4N\xi}}\ri)^N
\end{equation}
}

Alternatively, we can use the large $N$ limit to integrate out the
variables $z_2,...,z_N$ to get an effective theory for $z_1$
interacting with the droplet. The result is an integral for $z_1$
with effective action
$$
S_{\rm eff}=\frac{2N}{\lambda}\bar z_1 z_1-(N-1)\int
d^2z\ln|z_1-z|^2\rho(z)+\frac{k}{\sqrt{2}}(z_1+\bar z_1)$$ This
action has an electrostatic interpretation.  The fist term is the
potential due to a constant charge density on a large disc and it
tends to attract the oppositely charged particle with position $z_1$
to the center of the disc. The second is the repulsive
two-dimensional coulomb interaction with the charge distribution,
which, using (\ref{gaussiandensity}), is also a charged disc
centered on the origin. We expect some cancellation of the
attraction in the first term and the repulsion in the second term,
particularly in the interior of the discs.  The third term is the
contribution of an external electric field which pulls the $z_1$ to
the right in the complex plane. Using the constant density droplet,

\begin{eqnarray}
\int d^2w\rho(w)\log|w-z|^2 ~
=~\lf\{\matrix{{\small\frac{2}{\lambda}}|z|^2+\ln{\small\frac{\lambda}{2}}-1
&|z|<\sqrt{\frac{\lambda}{2}}\cr
 \ln |z|^2
  &|z|>\sqrt{\frac{\lambda}{2}}}\ri.
\label{Vsaddle}
\end{eqnarray}

we arrive at
\begin{equation}
S_{\rm eff}= \left\{ \matrix{ \frac{2}{\lambda}\bar z_1 z_1-(N-1)
\left(\ln\frac{\lambda}{2}-1\right) +\frac{k}{\sqrt{2}}(z_1+\bar
z_1)+{\cal O}(N^0) & |z_1|^2<\frac{\lambda}{2} \cr
\frac{2N}{\lambda}\bar z_1
z_1-(N-1)\ln|z_1|^2+\frac{k}{\sqrt{2}}(z_1+\bar z_1)+{\cal O}(N^0) &
|z_1|^2>\frac{\lambda}{2} \cr }\right. \label{effac}\end{equation}
Since we only found the eigenvalue density to leading order in large
$N$, only the order $N$ terms in this effective action should be
trusted. In the region $|z_1|^2>\frac{\lambda}{2}$, the effective
action has a minimum at its minimal value,
$|z_1|^2=\frac{\lambda}{2}$, and a steep slope, of order $N$ as
$|z_1|$ increases.  We conclude that, in the large $N$ limit, the
eigenvalue is confined to the disc $|z_1|<\frac{\lambda}{2}$. This
is just the same disc where the eigenvalue distribution
(\ref{gaussiandensity}) has support in the infinite $N$ limit.
Inside the disc, the terms of order $N$ in the energy are constant.
To that order, the potential there is flat and, to understand the
behavior of the particle, it is necessary to understand the terms of
order one. However, to correctly evaluate the order one terms in the
action (some of which have been written in (\ref{effac})) we would
need to properly evaluate the 1/N corrections to the density. If we
did so, we would find that the terms of order one are also constant
and the potential inside the disc is actually flat.  The average of
the position of the charge at $z_1$ over the degenerate saddle
point, which is the disc, and of course gives precisely the same
result as (\ref{case1}).

\item{}{\it Large winding loop}:
$\lim_{N\to\infty}{\small\frac{k}{N}}=\xi$, with $\xi$
a constant.

In ref.~\cite{Drukker}, it is argued that the loop
$W^{(k)}_{\tableau{1}}$ with large winding $k$ is dual to a D3-brane
whose world-volume is $AdS_2\times S^2$ embedded in $AdS_5$ and with
$k$ units of electric flux on the D3-brane. It is argued in
\cite{Gomis:2006sb} that the $AdS_2\times S^2$ D3-brane corresponds
to a Wilson loop in the symmetric representation
$W_{\tableau{2}\cdots\tableau{1}}$.

In the following, we will analyze the large winding number limit of
a single trace Wilson loop. In particular, the relation between
D3-brane and large winding loop is checked in the limit
\begin{equation}
k,N\riya\infty~~{\rm with}~~\ka={k\rt{\la}\o 4N}~~{\rm fixed}
\end{equation}

In this case, the saddle-point equation (\ref{saddpt}) still holds
and again the solution at leading order in large $N$ is the
disc-shaped droplet (\ref{gaussiandensity}). In the effective action
(\ref{effac}), there are now terms of order $N$ in the action inside
the disc as well as outside and it is these terms which must be
taken into account to get the leading order in large $N$. The
integral over $z_1$ has the form \begin{eqnarray}
\bra W^{(k)}_{\tableau{1}}\ket\sim
e^{-N\lambda\xi^2/8}\left( \int_{|z_1|^2<\frac{\lambda}{2}}
d^2z_1
e^{N\ln(\lambda/2)-N+N{\small\frac{\xi}{\sqrt{2}}}(z_1+\bar z_1)}
\right. \nonumber \\ \left. + \int_{|z_1|^2>\frac{\lambda}{2}}
d^2z_1 e^{
-{\small\frac{2N}{\lambda}}|z_1|^2+N\ln|z_1|^2+N{\small\frac{\xi}{\sqrt{2}}}(z_1+\bar
z_1)}\right)\label{fund10}
\end{eqnarray}

In the large $N$ limit, the classical position of the charge $z_1$
is outside of the disc.  The second term in (\ref{fund10}) dominates
and can be evaluated by saddle-point integration.  The saddle point
is
\begin{equation}
z_1^*=\rt{\la\o2}\,({\ka}+\rt{1+{\ka}^2})=
\rt{\la\o2}\,e^{2\sinh^{-1}{\ka}}
\label{zstar}
\end{equation}
The integral is
\begin{equation}
\bra W^{(k)}_{\tableau{1}}\ket~\sim~
e^{2N\left({\kappa}\sqrt{{\kappa}^2+1}+\sinh^{-1}{\kappa} \right)}
~~,~~{\kappa} = \frac{k\sqrt{\lambda}}{4N}
\label{nd}\end{equation} This expression agrees, as it should,
with one derived in ref.~\cite{Drukker} using the Hermitian matrix
model.  It was also shown in ref.~\cite{Drukker} that this agrees
with the Born-Infeld action of the D3-brane.  It was further
argued there that, for a large winding number Wilson loop, the
fundamental string world-sheet blows up into a D3-brane.

In our matrix model picture, the D3-brane corresponds to an isolated
eigenvalue residing outside of the droplet.  This has a strong
analogy with the bubbling geometry where a giant graviton, which is
also a D3-brane is also described in this way~\cite{Lin:2004nb}.

\item{}{\it intermediate case}: $k\sim N^\alpha$ with $0<\alpha<1$

In this case, the electric field due to the Wilson loop term in the
action is not strong enough to pull the charge out of the disc, but
in the limit $N\to\infty$ it is pulled to the edge where $z_1=\bar
z_1=\sqrt{\small\frac{\lambda}{2}}$.

\end{enumerate}

The essential difference that we have found between the small and
large winding number Wilson loop is that the eigenvalue which
interacts with the loop resides inside the droplet for small winding
number and outside of the droplet for large winding number.

\subsection*{Appendix A: An integral formula}\label{formula}

In the following, we shall prove the integral formula
\begin{equation}
\frac{
\int\prod_{i=1}^Ndx_ie^{-x_i^2}\lap(x)^2\prod_{i=1}^Ne^{k_ix_i}
}{\int\prod_{i=1}^Ndx_ie^{-x_i^2}\lap(x)^2}=\frac{\int\prod_{i=1}^Nd^2z_ie^{-|z_i|^2}|\lap(z)|^2\prod_{i=1}^N
e^{{k_i\o\rt{2}}(z_i+\b{z}_i)-\qu
k^2_i}}{\int\prod_{i=1}^Nd^2z_ie^{-|z_i|^2}|\lap(z)|^2}
\label{theformula}
\end{equation}
which relates the expectation value of a general exponential in the
Hermitian matrix model to one in the normal matrix model.

The matrix eigenvalue integral on the left-hand-side of
eq.~(\ref{theformula}) is written as
\begin{eqnarray}
\frac{
\int\prod_{i=1}^Ndx_ie^{-x_i^2}\lap(x)^2\prod_{i=1}^Ne^{k_ix_i}
}{\int\prod_{i=1}^Ndx_ie^{-x_i^2}\lap(x)^2} &=&{1\o
N!}\int\prod_{i=1}^N dx_i\Psi(x_1,\cdots, x_N)^2 \prod_i
\left(e^{k_ix_i}\right) \label{116}
\end{eqnarray}
where $\Psi(x_1,\cdots, x_N)$ is the Slater determinant which is the
wave-function for the $N$ fermions in a harmonic potential
\begin{eqnarray}
\Psi(x_1,\cdots, x_N)&=&\det_{1\leq i,j\leq N}(\psi_{i-1}(x_j)) \nn
\psi_n(x)&=&{1\o\rt{2^nn!\rt{\pi}}}H_n(x)e^{-\hf x^2} \label{psisea}
\end{eqnarray}
$H_n(x)=e^{x^2}(-\del_x)^ne^{-x^2}$ is the Hermite polynomial. In
(\ref{116}), we used the relation
\begin{equation}
{1\o N!}\Psi(x_1,\cdots,x_N)^2={1\o
Z_H}e^{-\sum_{i=1}^Nx_i^2}\lap(x)^2 \label{119}
\end{equation}
where $Z_H$ is the partition function of hermitian matrix model
\begin{equation}
Z_H=\int\prod_{i=1}^Ndx_ie^{-x_i^2}\lap(x)^2=2^{-{N^2\o2}}(2\pi)^{N\o2}
\prod_{n=1}^Nn!
\end{equation}
We can further rewrite (\ref{116}) in the operator formalism using
the $N$ free oscillators $a_i,a^\dag_i$
$([a_i,a^\dag_j]=\cob_{ij})$. Using the relation
$\h{x}_i={(a_i+a_i^\dag)\o\rt{2}}$, (\ref{116}) becomes
\begin{equation}
\frac{
\int\prod_{i=1}^Ndx_ie^{-x_i^2}\lap(x)^2\prod_{i=1}^Ne^{k_ix_i}
}{\int\prod_{i=1}^Ndx_ie^{-x_i^2}\lap(x)^2} ={1\o
N!}\bra\Psi|\prod_{i=1}^N e^{\hf k_i(a_i+a^\dag_i)}|\Psi\ket
\label{117}
\end{equation}
where $|\Psi\ket$ is the anti-symmetric $N$-particle state
\begin{eqnarray}
|\Psi\ket&=&|0\ket\wg|1\ket\wg\cdots\wg|N-1\ket \nn &=&\sum_{\si\in
S_N}(-1)^{{\rm deg}\si}|\si(1)-1\ket\tens\cdots\tens|\si(N)-1\ket
~~\in {\cal H}_1\tens\cdots\tens{\cal H}_N
\end{eqnarray}
and $|n\ket={1\o\rt{n!}}(a^\dag)^n|0\ket$ is the $n^{\rm th}$
excited state. Here ${\cal H}_i$ denotes the Fock space of $i^{\rm
th}$ oscillator and the operators $a_i,a_i^\dag$ act on ${\cal
H}_i$. The state $|\Psi\ket$ represents a Fermi sea of $N$ fermions
in a harmonic potential. To evaluate (\ref{117}), we need to know
the overlap $\bra n|e^{k\h{x}}|m\ket$. In order to rewrite the
hermitian matrix model to normal matrix model, the key step is to
insert the completeness relation of the coherent state
$|z\ket=e^{za^\dag}|0\ket$
\begin{equation}
{\bf 1}={1\o\pi}\int d^2z e^{-|z|^2}|z\ket\bra z|
\end{equation}
Then the matrix element $\bra n|e^{k\h{x}}|m\ket$ is written as
\begin{eqnarray}
\bra n|e^{k{a+a^\dag\o\rt{2}}}|m\ket&=&e^{-\qu k^2}\bra n|
e^{{k\o\rt{2}}a}e^{{k\o\rt{2}}a^\dag}|m\ket\nn &=&e^{-\qu
k^2}{1\o\pi}\int d^2ze^{-|z|^2}\bra n| e^{{k\o\rt{2}}a} |z\ket\bra
z|e^{{k\o\rt{2}}a^\dag}|m\ket\nn &=&{1\o\pi\rt{n!m!}} \int
d^2ze^{-|z|^2+{k\o\rt{2}}(z+\b{z})-\qu k^2}z^n\b{z}^m \label{118}
\end{eqnarray}
Let us consider a term appearing in (\ref{116}) of the form
\begin{equation}
\int\prod_{i=1}^Ndx_i\Psi(x_1,\cdots,x_N)^2\prod_{i=1}^Ne^{k_ix_i}
=\bra \Psi|\prod_{i=1}^Ne^{k_i\h{x}_i}|\Psi\ket
\end{equation}
Using the relation (\ref{118}), this is written as
\begin{eqnarray}
{1\o N!}\bra \Psi|\prod_{i=1}^Ne^{k_i\h{x}_i}|\Psi\ket&=& {1\o
N!}\sum_{\si,\tau\in
S_N}(-1)^{\si+\tau}\prod_{i=1}^N\bra\si(i)-1|e^{k_i\h{x}_i}|\tau(i)-1\ket
\nn &=&{1\o Z_N}\sum_{\si,\tau\in S_N}(-1)^{\si+\tau}
\prod_{i=1}^N\int d^2z_ie^{-|z_i|^2+{k_i\o\rt{2}}(z_i+\b{z}_i)-\qu
k^2_i} z^{\si(i)-1}_i\b{z}_i^{\tau(i)-1} \nn &=&{1\o
Z_N}\prod_{i=1}^N\int d^2z_i
e^{-|z_i|^2+{k_i\o\rt{2}}(z_i+\b{z}_i)-\qu k^2_i} \lap(z)\lap(\b{z})
\label{120}
\end{eqnarray}
where $Z_N$ is the partition function of normal matrix model
\begin{equation}
Z_N=\int\prod_{i=1}^Nd^2z_ie^{-|z_i|^2}|\lap(z)|^2=\pi^N\prod_{n=1}^Nn!
\end{equation}
From (\ref{119}) and (\ref{120}), we find the identity between the
Gaussian hermitian matrix model and the Gaussian normal matrix model
\begin{equation}
{1\o Z_{H}}
\int\prod_{i=1}^Ndx_ie^{-x_i^2}\lap(x)^2\prod_{i=1}^Ne^{k_ix_i}
={1\o
Z_{N}}\int\prod_{i=1}^Nd^2z_ie^{-|z_i|^2}|\lap(z)|^2\prod_{i=1}^N
e^{{k_i\o\rt{2}}(z_i+\b{z}_i)-\qu k^2_i} \label{121}
\end{equation}
We should stress that the relation (\ref{121}) is exact.

\section{Antisymmetric representations}

We showed in the previous section that, when the winding number $k$
of the loop was small, it was described by an eigenvalue moving in a
flat potential in the interior of the droplet. When $k$ was larger,
of order $N$, the particle moved outside of the droplet and is
interpreted as a D3-brane.  In both cases, the expectation value of
the Wilson loop was given by a certain electrostatic interaction
energy of the particle with the rest of the droplet.

In the dual string theory, the small winding loop is well described
by the boundary of fundamental string in $AdS_5\times S^5$. We will
now consider the situation when large number of such small winding
loops are put on top of each other. For definiteness, let us
consider the loop $\bra W_{A_K}\ket$ in the $K^{\rm th}$
anti-symmetric representation. In
refs.~\cite{Yamaguchi:2006tq,Gomis:2006sb}, it is argued that
$W_{A_K}$ is dual to a D5-brane of the shape $AdS_2\times S^4$ with
the $K$ unit of electric flux on it.

The $K^{\rm th}$ antisymmetric representation is the Young diagram
with one column of $K$ boxes,
 $$
 {\cal A}_K~=~\left. \tableau{1 1 1 1 1 1 1}\right\}~K~{\rm boxes}$$
 and we consider the expectation value of the Wilson loop in this representation,
 \begin{equation}\label{as1}
 \bra W_{{\cal A}_K}\ket =\frac{ \int [dM]
 e^{-{\small\frac{2N}{\lambda}}{\rm Tr}M^2}\frac{1}{{\rm dim}{\cal A}_K}{\rm
 Tr}_{{\cal A}_K}\left(e^M\right) }{ \int [dM]
 e^{-{\small\frac{2N}{\lambda}}{\rm Tr}M^2} }
 \end{equation}
 where ${\rm dim}{\cal A}_K=N!/K!(N-K)!$.
\footnote{For a general representation $R$ with Young tableaux
$(h_1,\cdots,h_N)$, the dimension is
$$
{\rm dim}R=\prod_{1\leq i<j\leq N}{h_i-h_j+j-i\o j-i}
$$
}
This can be written  as an integral over the
 eigenvalues of Hermitian matrices,
 \begin{equation}\label{as2}
 \bra W_{{\cal A}_K}\ket =(-1)^{K+1} \frac{ \int \prod_i dm_i
 e^{-{\small\frac{2N}{\lambda}}m_i^2} \Delta^2(m)\left(e^{ \sum_{i=1}^K m_i}\right) }{  \int \prod_i dm_i
 e^{-{\small\frac{2N}{\lambda}}m_i^2} \Delta^2(m) }
 \end{equation}
 Where we have used the Schur polynomial formula which gives us
 $$
 {\rm Tr}_{{\cal A}_K}\left({\rm diag}M\right)= (-1)^{K+1}\sum_{i_1<i_2<...<i_K} M_{i_1}M_{i_2}...M_{i_K}
 $$
 (See Eqn.~(\ref{112}) for the example of $K=2$.)
 Note that, in Eq.~(\ref{115}), the combinatorial factor that we would find from choosing
 the first $K$ eigenvalues cancels with the
 dimension of the representation.

 Once the Hermitian matrix integral is written in the form (\ref{as2}),
 we can apply the integral formula (\ref{theformula})
 to write the rewrite it as a normal matrix
model:\footnote{Note that there is an interesting $K\to N-K$
 duality in the integral (\ref{115}).  We find it by translating all
 of the eigenvalues $m_i\to m_i+\sqrt{\small\frac{\lambda}{8N}}$ and
 noting that the Vandermonde determinant is invariant under the
 simultaneous translation of all eigenvalues.  Then, we transform
 $m_i\to -m_i$ to obtain
 $$
 \bra W_{A_K}\ket= \bra W_{A_{N-K}}\ket\cdot  ~e^{\small \left( 2\frac{K}{N}-1\right)\frac{\lambda}{8}}
$$
In particular, $\bra W_{A_N}\ket=e^{\small\frac{\lambda}{8}}$. Since
$A_N$ is  trivial as a representation of $SU(N)$, $\bra W_{A_N}\ket$
is equal to the circle Wilson loop in the non-interacting adjoint
$U(1)$ gauge theory. }
 \begin{equation}\label{115}
 \bra W_{A_K}\ket =(-1)^{K+1}e^{-K\lambda/8N} \frac{ \int \prod_i d^2z_i
 e^{-{\small\frac{2N}{\lambda}}\bar z_i z_i} |\Delta(z)|^2\left(e^{{\small\sqrt{\frac{1}{2}}}
  \sum_1^K(z_i+\bar z_i)}\right) }
 {  \int \prod_i d^2z_i
 e^{-{\small\frac{2N}{\lambda}}\bar z_i z_i} |\Delta(z)|^2 }
 \end{equation}

To analyze this model, it is convenient to divide the eigenvalues
into two sets, those that appear in the loop $z_1,...,z_K$ and those
which do not, $z_{K+1},...,z_{N}$.

Let us first analyze the case where  $K$ remains finite as
$N\to\infty$, i.e. $\lim_{N\to\infty}\frac{K}{N}=0$.  In this case,
the set of $N-K$ eigenvalues can be integrated out by solving for
their saddle-point, which is again the circular droplet
(\ref{gaussiandensity}).  The remaining $K$ eigenvalues have an
effective action which concentrates them within the circular
droplet. Inside the droplet, the terms of order $N$ in the effective
action cancel, and to analyze the dynamics, one must take into
account the $1/N$ corrections to the density, to get the effective
action to order one.  A guess would be that the potential for
eigenvalues inside the drop is simply flat so they move freely
there.   Then, our result for infinite $N$, finite $K$ would be
\begin{equation}\label{as10}
\bra W_{A_K}\ket = (-1)^{K+1} \bra W_{\tableau{1}}\ket^K
\end{equation}
We shall see that this is borne out by a careful analysis below.

The effect of integrating out the  $N-K$ eigenvalues is summarized by
the joint probability distribution of remaining $K$ eigenvalues \cite{ginibre}
\begin{eqnarray}
\rho_N(z_1,\cdots,z_K)&=&\frac{N!}{(N-K)!} \frac{ \int
\prod_{a=K+1}^Nd^2z_a\,
e^{-{\small\frac{2N}{\lambda}}\sum_{i=1}^N|z_i|^2}\prod_{i<j}|z_i-z_j|^2}{\int
\prod_{i=1}^Nd^2z_a\,
e^{-{\small\frac{2N}{\lambda}}\sum_{i=1}^N|z_i|^2}\prod_{i<j
}|z_i-z_j|^2} \nn
&=&\left(\frac{2}{\pi\lambda}\right)^K~e^{-\sum_{i=1}^K{\small\frac{2N}{\lambda}}|z_i|^2}\det
\lf(\sum_{p=0}^{N-1} {({\small\frac{2N}{\lambda}}z_i\b{z_j})^p\o
p!}\ri) \label{joint}
\end{eqnarray}
 When $N\to\infty$, we must study the convergence of the sum in (\ref{joint}). It is easy to see that the single
 particle distribution converges to a positive constant if
 $|z_1|<\sqrt{\frac{\lambda}{2}}$ and zero if $|z_1|$ is outside of this
 radius~\cite{ginibre}. This means that the eigenvalues are confined
 to a droplet.  Then, it can be seen that the multi-particle
 distribution converges to a constant for all eigenvalues inside the
 droplet.
    In this way, we see that the eigenvalues are
confined to the droplet and have a flat potential when inside the
droplet.

 In the large $N$ limit with fixed $K$, $\rho$ becomes
\begin{equation}
\rho(z_1,\cdots,z_K)=\left({2\o\pi\lambda}\right)^K
e^{-\sum_{i=1}^K{\small\frac{2N}{\lambda}}|z_i|^2}\det(e^{{\small\frac{2N}{\lambda}}z_i\b{z}_j})
\to \left({2\o\pi\lambda}\right)^K~~~~
|z_i|<\sqrt{\frac{\lambda}{2}}
\end{equation}
 It is easy to see that this produces (\ref{as10}).  We will derive
 this result in a third way in the Appendix to this Section.

In the case where $K\to\infty$ and the ratio $K/N$ remains finite as
$N\to\infty$, we again divide the eigenvalues into two subsets, so
that one subset is those which couple directly to the Wilson loop
and the other is those which do not.  In this case, each subset
contains a finite fraction of the eigenvalues. We introduce two
densities,
\begin{equation}\label{twodensities}\rho(z)=\frac{1}{K}\sum_{i=1}^K\delta(z-z_k)~~,~~
\hat\rho(z) = \frac{1}{N-K}\sum_{i=K+1}^N\delta(z-z_i) ~~~\int
d^2z\rho(z)=1=\int d^2z\hat\rho(z)\end{equation} and the saddle
point equations are
$$
\frac{2}{\lambda}z=\frac{K}{N}\int d^2z'\frac{\rho(z')}{\bar z-\bar
z'}+ \frac{N-K}{N}\int d^2z'\frac{\hat\rho(z')}{\bar z-\bar
z'}+\frac{1}{\sqrt{2}N}~~~z\in{\rm support~of~}\rho
$$
$$
\frac{2}{\lambda}z=\frac{N-K}{N}\int d^2z'\frac{\hat\rho(z')}{\bar
z-\bar z'}+ \frac{K}{N}\int d^2z'\frac{\rho(z')}{\bar z-\bar
z'}~~~z\in{\rm support~of~}\hat\rho
$$
In the large $N$ limit, these would apparently be solved by any two
functions such that
\begin{equation}\label{splitdensity}
\frac{K}{N}\rho(z)+\frac{N-K}{N}\hat\rho(z) =\left\{ \matrix{
\frac{2}{\pi\lambda} & |z|<\sqrt{\frac{\lambda}{2}} \cr 0 &
|z|>\sqrt{\frac{\lambda}{2}} \cr } \right.
\end{equation}  They must add to
form the circular droplet density in Eq.~(\ref{gaussiandensity}),
but otherwise any split of the disc distribution into two functions
of the appropriate normalization will do. This is a highly
degenerate solution and, again, the degeneracy must be split by
corrections of order $1/N$. Here, it can be resolved by maximizing
the integral
$$E_{\rm int}= K\int d^2z\rho(z)(z+\bar z)$$ In the electrostatic
analogy, this term is the interaction of a subset of the charged
particles with an external electric field.  A fraction $K/N$ of the
particles interact with this field and the remaining fraction
$1-K/N$ do not. This polarizes the droplet by pushing the $K$
eigenvalues whose density is $\rho$ to the right of the droplet and
the other $N-K$ eigenvalues to the left (see figure 1b). The result
is
\begin{equation}\label{splitdensity1}
\rho(z)  = \frac{N}{K} \frac{2}{\pi\lambda} ~
\Theta\left({\small\sqrt{\small\frac{\lambda}{2}}}-|z|\right)\Theta(z+\bar
z-\sqrt{2\lambda}\cos\theta)
\end{equation}
\begin{equation}\label{splitdensity2}
 \hat\rho(z) = \frac{N}{N-K} \frac{2}{\pi\lambda} ~
\Theta\left({\small\sqrt{\small\frac{\lambda}{2}}}-|z|\right)\Theta(\sqrt{2\lambda}\cos\theta-z-\bar
z)
\end{equation}
where $\Theta$ is the Heavyside function and the normalization of
the densities in Eq.~(\ref{twodensities}) determines $\theta$:
$$
\theta-\frac{1}{2}\sin 2\theta=\pi\frac{K}{N}
$$
Then the expectation value of the Wilson loop is given by
\begin{equation}
\bra W_{A_K}\ket~=~\exp\left( \frac{1}{\sqrt{2}}\int
d^2z\rho(z)\left(z+\bar z\right)\right) =\exp\left(
\frac{2N\sqrt{\lambda}}{3\pi}\sin^3\theta\right)
\label{sy}\end{equation} which agrees with the result of Yamaguchi
\cite{Yamaguchi:2006tq} who argued that it matches the free energy
of the 5-brane computed using the Born-Infeld action and who
derived the same result from a Hermitian matrix model. This
agreement between hermitian model analysis and the normal model
can easily be seen directly by projecting the circular droplet to
the real axis of $z$. Then the eigenvalue density along the real
axis is given by the semi-circle law, which of course is a
eigenvalue distribution of the Gaussian hermitian matrix model.

We summarize the results so far in Figure 1.  For the small winding,
small representation loop, the eigenvalue which couples to the loop
lives in a flat potential inside the disc.  When the winding number
gets large, it acts like an electric field which pulls the
eigenvalue outside of the disc as depicted in fig. 1a. This is
associated with the D3-brane.  In the case where there is a large
multiplicity of small winding number loops, the eigenvalues still
reside in the disc, but the loop polarizes them as in fig. 1b. One
can think of the line which divides the two subsets of eigenvalues
in this case as the analog of a hole in the eigenvalue sea. In the
case of a chiral primary operator, such a hole is also interpreted
as a giant graviton which is extended in the $S^5$ direction.  In
the case of the Wilson loop, it seems to be more economical to
interpret it as a 5-brane~\cite{Gomis:2006sb}.

\begin{figure}[bht]
\centerline{\epsfxsize=3.5in\epsfbox{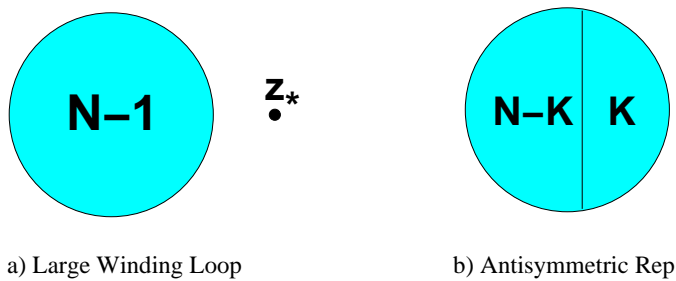}} \caption{\small
Eigenvalue distribution in the normal matrix representation of
Wilson loops. a) A large winding loop $W^{(k)}$ corresponds to a
single eigenvalue sitting at $z=z_1^*$ outside the droplet. b) For a
Wilson loop in the $K^{\rm th}$ antisymmetric representation $A_K$,
the droplet is divided by a line ${\rm Re}z=\rt{\la\o2}\cos\th$ and
$K$ eigenvalues and $N-K$ eigenvalues are distributed on the right
and left of this line, respectively.} \label{figure1}
\end{figure}

\subsection{Symmetric representation}

There seem to be two competing descriptions of the D3-brane.  In
ref.~\cite{Drukker} the claim was that the large wrapping number
single trace Wilson loop is dual to a D3-brane.  In
~\cite{Gomis:2006sb} the Wilson loop was one in a representation
consisting of a long single row Young diagram.  In the following
we will show that there are in fact two regimes.  If $\lambda $ is
large enough, the two descriptions are in fact identical in the
large $N$ limit.   If $\lambda$ is lowered beyond a certain
critical value $\sim 5.5$, there is a first order phase transition
to a regime where the Wilson loop behaves like a source of
D5-brane as described in
ref.~\cite{Yamaguchi:2006tq,Gomis:2006sb}.

We can also consider the Wilson loop in a representation which has
Young diagram a single row $\tableau{3}...\tableau{1}$ with $k$
boxes. The decomposition of the character in this representation
into symmetric polynomials, for a diagonal matrix is
\begin{equation}\label{symm1}
{\rm Tr}_{\tableau{3}...\tableau{1}}e^M =\frac{K!(N-1)!}{(N+K-1)!}
\sum_{i_1\leq i_j\ldots\leq i_k}e^{m_{i_1}+m_{i_2}+...+m_{i_k}}
\end{equation}
In order to use the formula (\ref{theformula}), it is convenient
to divide the sum in Eq.~(\ref{symm1}) into special cases. Here,
we will compare the two extreme such cases, first the set of terms
where all of the indices are the same. This  corresponds to a
single large winding Wilson loop. Second, we will consider the
case where all indices are different. This gives a contribution
identical to the antisymmetric representation Wilson loop. In both
cases, we will consider the case where $k$ is large.

One would naively expect that the case with all indices different
would dominate simply because it has a much higher multiplicity.
In fact, if we were averaging the right-hand-side of (\ref{symm1})
under an integral which was symmetric in the indices, the terms
where all indices are different could be replaces by
$$
 \frac{N!}{k!(N-k)!}e^{m_{1}+m_{2}+...+m_{k}}
$$
The factor comes from the multiplicity of these terms.

The other term of interest would be replaced by
$$
Ne^{km_1}
$$
and it occurs which a smaller factor.

 The expectation value of the first of these is identical to the
 expectation value of the Wilson loop in an anti-symmetric
 representation which we have already computed.  Taking into
 account the multiplicity factor and using Eq.~(\ref{sy}) we get

\begin{equation}
 \exp\left(N\left(-
 {\small\frac{k}{N}}\ln{\small\frac{k}{N}}-(1-{\small\frac{k}{N}})\ln(1-{\small\frac{k}{N}})+
\frac{2\sqrt{\lambda}}{3\pi}\sin^3\theta\right)\right)
\label{sy1}\end{equation} where $ \theta-\frac{1}{2}\sin
2\theta=\pi\frac{k}{N} $.

This should be compared with the exponential in N part of the
expectation value of $ Ne^{km_1} $ which we can read off from
Eq.~(\ref{nd}), which gives
\begin{equation}
\exp\left(N\left(2{\kappa}\sqrt{{\kappa}^2+1}+2\sinh^{-1}{\kappa}
\right)\right) ~~,~~{\kappa} = \frac{k\sqrt{\lambda}}{4N}
\label{nd1}\end{equation} To compare these, we choose
${\small\frac{k}{N}}={\small\frac{1}{2}}$ and plot the difference
between the exponents in Eqs.~(\ref{sy1}) and (\ref{nd1}) as a
function of $\lambda$.  The plot is depicted in figure 2.  There
we see a crossover between two regimes: when $\lambda< 5.5$ the
case (\ref{sy1}) dominates whereas for $\lambda > 5.5$ the case
(\ref{nd1}) dominates.  We have not analyzed intermediate cases
where sum but not all eignevalues are equal.  We speculate that
there is a phase transition between these two regimes.  Seen from
the point of view of the electrostatic analogy, in the first phase
the eigenvalues which couple to the external electric field prefer
to be independent.  In the second phase they prefer to sit on top
of each other as one macroscopically charged particle. This occurs
when $\lambda$ is sufficiently large.
\begin{figure}[htb]
\centerline{\epsfxsize=3.5in\epsfbox{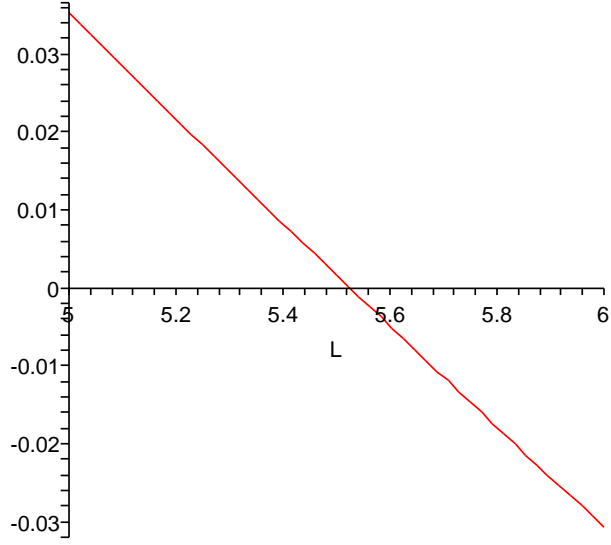}}
\caption{\small This is the graph of the differences in the
exponents in (\ref{sy1}) minus (\ref{nd1}) (divided by $N$) in the
region where they cross.  There is a first order phase transition
where $\lambda\approx 5.5$.  This figure is plotted for the case
where $k=N/2$.} \label{figure2}
\end{figure}

\subsection*{Appendix B: Contour Integral Representation}
In this Appendix, we rederive
(\ref{as10}) by directly integrating out the $N-K$ eigenvalues
using the joint probability distribution (\ref{joint}).
\begin{equation}
\bra W_{A_K}\ket={N!\o(N-K)!}e^{-{K\la\o8N}}
\prod_{i=1}^K\int{d^2z_i\o\pi}e^{-|z_i|^2+\hf\rt{\la\o N}(z_i+\b{z}_i)}
\det\lf(\sum_{p=0}^N{(z_i\b{z}_j)^p\o p!}\ri)
\end{equation}
Here and the following section, we use the rescaled variables
$z_i^{\rm new}=\rt{2N\o\la}z_i^{\rm old}$ to make
the Gaussian part $|z_i^{\rm new}|^2$ simple.

As we will explain in the next subsection of this Appendix,
we can integrate out the radial coordinate of $z_i$
and the the angular integral reduces to a contour
integral.
\begin{eqnarray}
&&\bra W_{A_K}\ket \nn &=&{N!\o(N-K)!}e^{-{K\la\o8N}}
\prod_{i=1}^K\int{d^2z_i\o\pi}e^{-|z_i|^2+\hf\rt{\la\o N}(z_i+\b{z}_i)}
\sum_{p_1,\cdots,p_K=0}^{N-1}\sum_{\si\in S_K}
(-1)^\si\prod_{i=1}^K{z_i^{p_{\si(i)}}\b{z_i}^{p_i}\o p_i!} \nn
&=&{N!\o(N-K)!}e^{{K\la\o8N}}\prod_{i=1}^K\oint{dz_i\o2\pi i z_i}
e^{\hf\rt{\la\o N}z_i}\sum_{p_1,\cdots,p_K=0}^{N-1}\sum_{\si\in S_K}
(-1)^\si\prod_{i=1}^K{(z_i+\hf\rt{\la\o N})^{p_{\si(i)}}\o
z_i^{p_i}} \nn &=&
{N!\o(N-K)!}e^{{K\la\o8N}}\prod_{i=1}^K\oint{dz_i\o2\pi i z_i}
e^{\hf\rt{\la}z_i}\sum_{p_1,\cdots, p_K=0}^{N-1}\sum_{\si\in
S_K}(-1)^\si \prod_{i=1}^K(z_{\si(i)}+{\la\o 2N})^{p_i}z_i^{-p_i}
\nn &=&
{N!\o(N-K)!}e^{{K\la\o8N}}\prod_{i=1}^K\oint{dz_i\o2\pi
i}e^{\hf\rt{\la}z_i} \det_{1\leq i,j\leq
K}\lf({(z_i+{\rt{\la}\o2N})^Nz_j^{-N}-1\o z_i-z_j
+{\rt{\la}\o2N}}\ri)
\label{ointAK}
\end{eqnarray}
We should stress that this is an exact expression for
 anti-symmetric representation at finite $N$ and $K$.
In the large $N$ limit with finite $K$,
the dominant contribution comes from the diagonal elements
of the matrix inside the determinant.
In this way, we see that the integral factorizes
\begin{eqnarray}
\bra
W_{A_K}\ket&\sim&
{N!N^K\o(N-K)!}\lf({2\o\rt{\la}}\ri)^K\prod_{i=1}^K\oint{dz_i\o2\pi i}
e^{\hf\rt{\la}z_i}\lf[\lf(1+{\rt{\la}\o2Nz_i}\ri)^N-1\ri] \nn
&\sim&\lf({2\o\rt{\la}}\ri)^K\prod_{i=1}^K\oint{dz_i\o2\pi
i}e^{\hf\rt{\la}(z_i+z_i^{-1})}=\lf({2\o\rt{\la}}I_1(\rt{\la})\ri)^K=
\bra W_{\tableau{1}}\ket^K \label{factorization}
\end{eqnarray}
Although (\ref{ointAK}) is an exact formula, it seems difficult
to take the large $N$ and $K$ limit with fixed $K/N$ in this expression.

\subsubsection*{Integrating Out the Radial Coordinate $|z|$
in (\ref{theformula})}
As promised above, let us
consider the integration of the radial coordinates $|z_i|$ of the
normal matrix eigenvalues $z_i$.
The remaining angular integral can be written as a contour integral.

In (\ref{118}), the matrix element  of $e^{{k\o\rt{2}}(a+a^\dag)}$
was written as the integral over $z$ in the coherent state $|z\ket$.
We write this as an integral over $|z|$ and ${\rm arg}(z)=\th$
\begin{eqnarray}
\bra n|e^{k\h{x}}|m\ket&=&{1\o\rt{n!m!}}\int_0^\infty
d|z|^2\int_0^{2\pi}{d\th\o2\pi}
e^{-|z|^2+{k\o\rt{2}}|z|(e^{i\th}+e^{-i\th})-\qu k^2}
|z|^{n+m}e^{i\th(n-m)} \nn
&=&{1\o\rt{n!m!}}\int_0^\infty dt\oint{dz\o2\pi iz}
e^{-t+{k\o\rt{2}}\rt{t}(z+z^{-1})-\qu k^2}t^{n+m\o2}z^{n-m}
\end{eqnarray}
where we introduced the variable $t=|z|^2$ and rewrote the $\th$-integral as
a contour integral.
After rescaling $z\riya t^{-\hf}z$, the $t$-integral can be performed
easily
\begin{eqnarray}
\bra n|e^{k\h{x}}|m\ket&=&{1\o\rt{n!m!}}\int_0^\infty dt\oint{dz\o2\pi iz}
e^{-t+{k\o\rt{2}}z+{k\o\rt{2}}tz^{-1}-\qu k^2}t^{m}z^{n-m} \nn
&=&\rt{m!\o n!}\oint{dz\o2\pi i}e^{{k\o\rt{2}}z-\qu k^2}
{z^m\o(z-{k\o\rt{2}})^{m+1}} \nn
&=&\rt{m!\o n!}\oint{dz\o2\pi i}e^{{k\o\rt{2}}z+\qu k^2}
{(z+{k\o\rt{2}})^n\o z^{m+1}}
\end{eqnarray}
In the last step, we have shifted $z\riya z+{k\o\rt{2}}$.
Using this expression, our key formula (\ref{120}) is written as
a contour integral
\begin{eqnarray}
&&{1\o N!}\bra\Psi|\prod_{i=1}^Ne^{k_i\h{x_i}}|\Psi\ket\nn
&=&{1\o N!}\sum_{\si,\tau\in S_N}(-1)^{\si+\tau}\prod_{a=1}^N
\oint{dz_a\o2\pi iz_a}e^{{k_a\o\rt{2}}z_a+\qu k_a^2}
{(z_a+{k_a\o\rt{z}})^{\si(a)-1}\o z_a^{\tau(a)-1}} \nn
&=&{1\o N!}\prod_{a=1}^N\oint{dz_a\o2\pi iz_a}e^{{k_a\o\rt{2}}z_a+\qu k_a^2}
\prod_{1\leq i<j\leq N}(z_i+{k_i\o\rt{2}}-z_j-{k_j\o\rt{2}})
(z^{-1}_i-z^{-1}_j)
\label{psicont}
\end{eqnarray}
This almost looks like a unitary matrix model.
For example, when
\begin{equation}
k_1=\cdots=k_K=\ell,\quad k_{K+1}=\cdots=k_K=0
\end{equation}
(\ref{psicont}) becomes a coupled system of $U(K)$ and $U(N-K)$ unitary
matrix model
\begin{eqnarray}
{1\o N!}\bra\Psi|\prod_{i=1}^Ke^{\ell \h{x}_i}|\Psi\ket
&=&{1\o N!}e^{\qu K\ell^2}\prod_{i=1}^K\int_0^{2\pi}{d\th_i\o2\pi}
e^{{\ell\o\rt{2}}e^{i\th_i}}|\lap(\th)|^2
\prod_{j=1}^{N-K}\int_0^{2\pi}{d\phi_j\o2\pi}|\lap(\phi)|^2 \nn
&&\times\prod_{i=1}^K\prod_{j=1}^{N-K}
(e^{i\th_i}+{\ell\o\rt{2}}-e^{i\phi_j})(e^{-i\th_i}-e^{-i\phi_j})
\end{eqnarray}

\section{OPE of Wilson Loops and Chiral Primary Operators}

So far the normal matrix model representation of Wilson loop and the
droplet picture which emerges at large $N$ are interesting technical
tools, but all of the computations that we have outlined could also
have been done, in principle, using the Hermitian Gaussian matrix
model.  However, when we consider the interaction between Wilson
loop and chiral primary operator, the normal matrix model
representation provides us with new insight.

As discussed in ref.~\cite{Lin:2004nb}, a 1/2-BPS chiral primary
operator has a dual description in supergravity which is specified
by the shape of a droplet. If we think of the chiral primary
operator as a state, the characteristic of that state which is
visible in geometry is the shape of the droplet that is associated
with it.

The disc-shaped droplet that we have discussed so far is the ground
state. The state of IIB string theory corresponding to it is the
ground state on the $AdS_5\times S^5$ background.  Small chiral
primaries correspond to the same background with some gravitons
excited.  Larger chiral primaries are giant gravitons or, if they
are large enough, they modify the geometry and the state is string
theory on a different background. In the large $N$ limit these
states are all characterized by droplets which, now are different
from the ground state disc droplet.

The vacuum expectation value of the Wilson loop operator measures
the vacuum matrix element and gives the content of the unit operator
in the loop.  We can consider some other matrix elements as well,
such as the matrix element between the vacuum and the state which is
created by a chiral primary operator.  This matrix element is of
course given in the integral in eq.~\ref{mmcp}.  In the language of
eigenvalues, it measures the ability of the Wilson loop operator to
modify the circular droplet which characterizes the ground state so
that it fits with the droplet for an excited state, the chiral
primary.



Let us consider the OPE between the Wilson loop and the chiral
primary operator ${\cal O}_k={1\o \rt{kN^k}}\Tr Z^k$. In the next
Section, we shall prove that it is given by a normal matrix model
\begin{equation}
\bra W_{\tableau{1}}{\cal O}_k\ket= 2^{-{\small\frac{k}{2}}}{1\o
Z_N}\int_{[z,\b{z}]=0} [d^2z]e^{-\Tr(z\b{z})} {1\o N}\Tr
\,e^{\hf\rt{\la\o N}(z+\b{z})-{\la\o8N}}{1\o \rt{kN^k}}\Tr z^k
\label{normOPE}
\end{equation}
The normalization of ${\cal O}_k$ is chosen so that the two-point
function is normalized to unity in the large $N$ limit. In the
eigenvalue basis, (\ref{normOPE}) is written as
\begin{equation}
\bra W_{\tableau{1}}{\cal
O}_k\ket=2^{-{\small\frac{k}{2}}}{e^{-{\la\o8N}}\o Z_N} \int
\prod_{i=1}^Nd^2z_ie^{-|z_i|^2}|\lap(z)|^2{1\o
N\rt{kN^k}}\sum_{i,j=1}^N e^{\hf\rt{\la\o N}(z_i+\b{z}_j)}z_j^k
\end{equation}

This is evaluated exactly using the joint probability distribution
(\ref{joint}).\footnote{This is a more subtle application of the
joint probability distribution since it probes its next-to-leading
order behavior in $1/N$. The naive leading order in large $N$,
which would be the factorized product of two traces, vanishes
after the integral over phases of $z$. This means that we must
take into account the next-to-leading order.  We do this by
integrating the exact formula to get (\ref{WOkresult}) and then
taking the large $N$ limit.}

 As discussed in the previous section, we
can integrate out the radial coordinate $|z_i|$ and rewrite this OPE
coefficient as a contour integral. However, to see the relation to
the droplet picture, we find it useful to integrate out the angular
part of $z_i$. After some computation, we find the exact finite $N$
expression for the normal matrix integral (\ref{normOPE})
\begin{equation}
\bra W_{\tableau{1}}{\cal O}_k\ket={e^{-{\la\o8N}}\o N\rt{kN^k}}\int_0^\infty d|z|^2\,e^{-|z|^2}
\sum_{i=1}^{k}
{|z|^{2(N-i)}\o(N-i)!}|z|^kI_k\lf(\rt{\la \o N}|z|\ri)
\label{WOkinI}
\end{equation}
In the large $N$ limit, the $|z|$-integral is localized to
the saddle point $|z|=\rt{N}$.
Physically, this corresponds to the localization of the eigenvalue to
the boundary of circular droplet.
This is consistent with the picture that the chiral primary ${\cal O}_k$
with $k\ll N$ corresponds to a graviton which in turn is represented by
a small ripple of the Fermi surface.
Finally, the OPE coefficient at large $N$ is found to be
\begin{equation}
\bra W_{\tableau{1}}{\cal O}_k\ket={1\o
N}2^{-{\small\frac{k}{2}}}\rt{k}I_k(\rt{\la}) \label{WOkresult}
\end{equation}
which produces the expression in Eq.~(\ref{ladsum}) which was
originally derived in ref.~\cite{Semenoff:2001xp}.

In the derivation of (\ref{WOkresult}) we have assumed that $k\ll
N$.  One interesting problem where this is not the case is the BMN
limit $k=J\sim\rt{N}$.  In this limit, the integral (\ref{WOkinI})
behaves differently. To take the BMN limit, it is useful to
rewrite the integral (\ref{WOkinI}) as a sum of Laguerre
polynomials
\begin{equation}
\bra W_{\tableau{1}}{\cal O}_k\ket={1\o N\rt{kN^k}}
\lf(\la\o 4N\ri)^{k\o2}e^{\la\o 8N}
\sum_{i=1}^kL_{N-i}^k\lf(-{\la\o 4N}\ri)
\label{Lagsum}
\end{equation}
Using the contour integral representation of Laguerre polynomials
\begin{equation}
a^kL^k_m(-a^2)=\oint{dw\o2\pi i}e^{aw}w^{k-1}(1+aw^{-1})^{m+k}
\end{equation}
the summation over $i$ in (\ref{Lagsum}) can be easily performed.
In this way, we arrive at the contour integral
representation of the normal matrix integral (\ref{normOPE})
\begin{equation}
\bra W_{\tableau{1}}{\cal O}_J\ket={2\o\rt{J\la}}e^{\la\o8N}\oint{dw\o2\pi i}
w^Je^{{\rt{\la}\o2}w}\lf(1+{\rt{\la}\o2Nw}\ri)^N
\lf[\lf(1+{\rt{\la}\o2Nw}\ri)^J-1\ri]
\label{contourWO}
\end{equation}
In this expression, it is easy to take the BMN limit
\begin{equation}
N,J\riya\infty\quad g_2={J^2\o N},\la'={\la\o J^2}~~{\rm fixed}
\end{equation}
In this limit the integral (\ref{contourWO}) becomes
a non-trivial function of $g_2$ and $\la'$
\begin{equation}
\bra W_{\tableau{1}}{\cal O}_J\ket\sim{2\o \rt{J^3\la'}}e^{g_2\la'\o8}\oint{dw\o2\pi i}
w^Je^{{J\rt{\la'}\o2}(w+w^{-1})}\Big(e^{g_2\rt{\la'}\o2w}-1\Big) \nn
\end{equation}
This integral is evaluated around the saddle point of
the term proportional to $J$
\begin{equation}
\del_w \lf[{\rt{\la'}\o2}(w+w^{-1})+\log w\ri]=0~~
\longrightarrow w_*=\rt{1+{1\o\la'}}-{1\o\rt{\la'}}
\end{equation}
Then we find
\begin{equation}
\bra W_{\tableau{1}}{\cal O}_J\ket={1\o N} \,{\cal A}\,e^{\cal S}
\end{equation}
where ${\cal S}$ and ${\cal A}$ are given by
\begin{eqnarray}
{\cal S}&=&\rt{\la}\lf[\rt{1+{1\o\la'}}+{1\o\rt{\la'}}\log\lf(\rt{1+{1\o\la'}}
-{1\o\rt{\la'}}\ri)\ri] \nn
{\cal A}&=&{1\o\rt{2\pi}}(1+\la'^{-1})^{-\qu}e^{{g_2\o8}(1+\rt{1+\la'})^2}\,
{\sinh\lf[{g_2\o4}(1+\rt{1+\la'})\ri]
\o{g_2\o4}(1+\rt{1+\la'})}
\end{eqnarray}
Since $g_2$ is the genus-counting parameter in the BMN limit,
this expression gives
an all-genus result of OPE between a single winding
Wilson loop $W_{\tableau{1}}$
and the 1/2 BPS BMN operator $\Tr Z^J$.
The expression of
${\cal S}$ agrees with the result of \cite{Zarembo:2002ph}.

On the other hand, the coupling to the large winding loop $W^{(w)}$ ($w\gg1$)
behaves very differently. As we saw above, the saddle point of
$z$-integral is at $z=z_1^*$ (\ref{zstar}).
Therefore, the OPE coefficient
behaves as
\begin{equation}
\bra W_{\tableau{1}}^{(w)}{\cal O}_k\ket\sim e^{2N\ka\rt{\ka^2+1}+2(N+k)\sinh^{-1}\ka}
\end{equation}

We can write down the OPE between the Wilson loop in the
$K^{\rm th}$ anti-symmetric representation and the chiral primary ${\cal O}_J$
in the form of the contour integral as before
\begin{equation}
\bra W_{A_K}{\cal O}_J\ket={KN!\o (N-K)!\rt{J}}\oint\prod_{i=1}^K{dz_i\o2\pi i}
e^{{\rt{\la}\o2}z_i}\lf(z_1+{\rt{\la}\o2N}\ri)^J\det A
\end{equation}
where $K\times K$ matrix $A$ is given by
\begin{eqnarray}
A_{i1}&=&{(z_i+{\rt{\la}\o2N})^{N-J}\o z_1^N}\cdot
{(z_i+{\rt{\la}\o2N})^{J}-z_1^J\o z_i+{\rt{\la}\o2N}-z_1} \nn
A_{ij}&=&{(z_i+{\rt{\la}\o2N})^Nz_j^{-N}-1\o z_i+{\rt{\la}\o2N}-z_j} \quad
(j\not=1)
\end{eqnarray}
When $K$ is of order $1$ and $N$ is sent to infinity,
the dominant contribution comes from
the diagonal elements of the matrix $A$
\begin{eqnarray}
\bra W_{A_K}{\cal O}_J\ket&=&{K\o N^K\rt{J}}
\oint\prod_{i=1}^K{dz_i\o2\pi i}
e^{{\rt{\la}\o2}z_i}\lf(z_1+{\rt{\la}\o2N}\ri)^J
\prod_{i=1}^KA_{ii} \nn
&\sim&K\lf({2\o\rt{\la}}I_1(\rt{\la})\ri)^{K-1}{1\o N}\rt{J}I_J(\rt{\la}) \nn
&=&K\bra W_{\tableau{1}}\ket^{K-1}\bra W_{\tableau{1}}{\cal O}_J\ket
\end{eqnarray}
Namely, the OPE coefficient factorizes when $K$ is small.

The more natural basis of chiral primary is given by the
Schur polynomial of $Z={\rm diag}(z_1,\cdots,z_N)$
\begin{equation}
{\cal O}_R={\det(z_i^{h_j+N-j})\o\det(z_i^{N-j})}={\det(z_i^{h_j+N-j})
\o\lap(z)}
\end{equation}
Here $h_1\geq h_2\cdots\geq h_N$ are the row length of the Young diagram
$R$. Using the normal matrix model, one can easily show that
the two-point function of ${\cal O}_R$ is diagonal in $R$.
Similarly, one can show that the OPE between $W_{\tableau{1}}$
and the chiral primary ${\cal O}_R$ is non-zero only if $R$ is a single
hook {\it i.e.,} $h_1\geq 1, h_2=\cdots=h_k=1,h_{k+1}=\cdots=h_N=0$
\begin{equation}
\bra W_{\tableau{1}}{\cal O}_R\ket=0\quad {\rm if}~~~R\not=\tableau{8 1 1 1 1}
\end{equation}
Explicitly, the OPE is given by
\begin{eqnarray}
\bra W_{\tableau{1}}
{\cal O}_{\tableau{4 1 1 1}}\ket&=&
{(-1)^{k-1}\o N}N^{|R|\o2}
\oint{dz\o2\pi iz}e^{{\rt{\la}\o2}z}z^{|R|}
\lf(1+{\rt{\la}\o2Nz}\ri)^{N+h_1-1} \nn
&\sim&{(-1)^{k-1}\o N}(N+h_1)^{|R|\o2}I_{|R|}\Big(\rt{g_{\rm YM}^2(N+h_1)}
\Big)
\end{eqnarray}
where $k$ is the length of the first column
and $|R|=h_1+k-1$ is the number of boxes.
It is interesting that $N$ is effectively replaced by $N+h_1$.

For a more general Wilson loop $W_R$,
the multi-hook chiral primary can couple to it.
The computation becomes harder as the number of hooks increases.
Therefore, we restrict ourselves to the case of
single-hook chiral primary.
The OPE between Wilson loop in
the $K^{\rm th}$ anti-symmetric representation
and the single-hook chiral primary is easily obtained as
\begin{equation}
\bra W_{A_K}{\cal O}_{\tableau{4 1 1 1}}\ket
=\oint \prod_{i=1}^K{dz_i\o2\pi i}e^{{\rt{\la}\o2}z_i}\,z_1^{-1}\lf(z_1
+{\rt{\la}\o2N}\ri)^{h_1+k-1}\det B
\end{equation}
where $K\times K$ matrix $B$ is given by
\begin{eqnarray}
B_{i1}&=&\lf(z_i+{\rt{\la}\o2N}\ri)^{N-k} z_1^{-N+k} \nn
B_{ij}&=&{(z_i+{\rt{\la}\o2N})^Nz_j^{-N}-1\o z_i+{\rt{\la}\o2N}-z_j}
\quad(j\not=1)
\end{eqnarray}
Again, when $K$ is of order $1$ the dominant contribution comes from
the diagonal part of $B$. In this case one finds
\begin{equation}
\bra W_{A_K}{\cal O}_{\tableau{4 1 1 1}}\ket
\sim K\bra W_{\tableau{1}}\ket^{K-1}\bra  W_{\tableau{1}}{\cal O}_{\tableau{4 1 1 1}}\ket
\end{equation}
As in the case of the Wilson loop correlator $\bra W_{A_K}\ket$,
we expect that the OPE coefficient $\bra W_{A_K}{\cal O}_R\ket$
fails to factorize when $K\sim N$. We leave this analysis as an
interesting future problem.

\subsection*{Appendix C:
Normal Matrix Representation of OPE between Wilson Loop and Chiral Primary}
The OPE between Wilson loop and
chiral primary is given by a complex matrix model
\begin{equation}
\int [d^2z]e^{-\Tr(z\b{z})}{1\o N}\Tr \,e^{\hf(z+\b{z})}{\cal
O}_R(\b{z}) \label{compOPE}
\end{equation}
Here, for notational simplicity, we have set ${\la\o 2N}=1$. This
parameter can easily be restored by the rescaling of the matrix
variable. Eq.~(\ref{compOPE}) can also written as integrals over the
real and imaginary parts of the complex matrix $z$,
\begin{equation}
\int [dXdY]e^{-\Tr(X^2+Y^2)}{1\o N}\Tr \,e^X {\cal O}_R(X-iY)
\end{equation}
Then we can formally do the integral over the imaginary part $Y$ to
get
\begin{eqnarray}
&&\int [dXdY]e^{-\Tr(X^2+Y^2)}{1\o N}\Tr \,e^X\,e^{-i\Tr Y^T{\del\o\del X}}
{\cal O}_R(X) \nn
&=&\int [dX]e^{-\Tr X^2}{1\o N}\Tr \,e^X\, e^{-\qu\Tr\lf({\del\o\del X}
\ri)^2}{\cal O}_R(X)
\end{eqnarray}
Note that we have taken into account the Wick contractions of powers
of $Y$ which had appeared in the chiral primary by operating with a
derivative operator.   Next we can diagonalize the Hermitian matrix
$X$ by conjugating it with a unitary matrix $U$
\begin{equation}
X=U{\rm diag}(x_1,\cdots,x_N)U^{-1}
\end{equation}
In the eigenvalue basis, the matrix Laplacian
is given by
\begin{equation}
\Tr\lf({\del\o\del X}
\ri)^2={1\o\lap(x)}\sum_{i=1}^N{\del^2\o\del x_i^2}\lap(x)+\sum_{i\not=j}{1\o(x_i-x_j)^2}{\del^2\o\del \om_{ij}\del\om_{ji}}
\end{equation}
where $\om=U^{-1}dU$ is the ``angular'' part. Since ${\cal O}_R(X)$
is independent of $U$, the Calogero-type term in the matrix
Laplacian does not contribute. Therefore, the matrix integral
reduces to the eigenvalue model
\begin{equation}
\int \prod_{i=1}^Ndx_ie^{-x_i^2}\lap(x)^2{1\o N}\sum_{i=1}^Ne^{x_i}
{1\o\lap(x)}e^{-\qu\sum_{i=1}^N\del_i^2}\lap(x){\cal O}_R(x)
\label{eigenx}
\end{equation}
for real eigenvalues $x_i$.   Now, using the identity for an
arbitrary function $f(x)$
\begin{equation}
e^{-\qu\del_x^2}f(x)=e^{\hf x^2}f\lf({x-\del_x\o2}\ri)e^{-\hf x^2}
\end{equation}
the integral (\ref{eigenx}) becomes
\begin{equation}
\int \prod_{i=1}^Ndx_i \,e^{-\hf\sum_ix_i^2}\lap(x){1\o N}\sum_{i=1}^Ne^{x_i}
\,{\cal O}_R\lf({x-\del_x\o2}\ri)\lap\lf({x-\del_x\o2}\ri)e^{-\hf\sum_ix_i^2}
\label{laponvac}
\end{equation}
Now, we notice that the combination ${x_i-\del_i\o\rt{2}}=a_i^\dag$
is a creation operator and $e^{-\hf\sum_ix_i^2}$ is the wavefunction
of the harmonic oscillator vacuum. The first factor and the last
factor in(\ref{laponvac}) are both Slater determinants of the ground
state wave-function of a set of $N$ fermions moving in a harmonic
oscillator potential, see eq.~(\ref{psisea}), up to a normalization
factor
\begin{equation}
{\cal N}\lap(x)e^{-\hf\sum_ix_i^2}
={\cal N'}\lap\lf({x-\del_x\o2}\ri)e^{-\hf\sum_ix_i^2}
=\Psi(x_1,\cdots,x_N)
\end{equation}
Using the bracket notation as in section 2.4,
the integral is written as
\begin{equation}
\bra\Psi|{1\o N}\sum_{i=1}^Ne^{{1\o\rt{2}}(a_i+a_i^\dag)}{\cal O}_R(a^\dag)
|\Psi\ket
\label{sandSR}
\end{equation}
Finally, inserting the completeness relation for coherent states,
(\ref{sandSR}) becomes a normal matrix model in the same way as we
saw in section 2.4.

This derivation of the normal matrix model shows that this relation
is valid for any chiral primary ${\cal O}_R$. On the other hand, the
Wilson loop part contains a normal ordering constant in general,
hence the equivalence between complex matrix model and normal matrix
model holds only  for a certain representation where this
normal ordering factor is common to all terms in the representation.
For instance, the normal matrix model is exact for the OPE
computation of the Wilson loop in the anti-symmetric representation.

\section{Discussion}

In this paper, we have shown  that the expectation values of
circular Wilson loops and the OPE between Wilson loops and chiral
primaries both can be written as normal matrix models. This
representation as a normal matrix model is exact for some
representations of Wilson loop and it is approximate for some
other cases.

This normal matrix model gives a unified description of 1/2-BPS
chiral primary operators and the 1/2-BPS Wilson loops. In
particular, there is a nice analogy between dual giant graviton
and large winding Wilson loop. Namely, the dual giant graviton
wrapped around $S^3$ inside $AdS_5$ is identified as a particle
outside the droplet, whereas the large winding Wilson loop also
corresponds to an isolated eigenvalue outside the droplet.
Moreover it has been proposed that the object dual to a large
winding loop is a D3-brane with world-volume $AdS_2\times S^2$ and
extended into $AdS_5$.  The direction that the dual object blows
up into ({\it i.e.}, $AdS_5$ direction) is the same as the case of
the giant graviton.

In the case of giant graviton wrapped around an $S^3$ inside
$S^5$, the droplet picture is a hole inside the droplet and the
corresponding operator is a Schur polynomial of anti-symmetric
representation. The corresponding Wilson loop is in an
anti-symmetric representation of gauge group, and the dual object
is a D5-brane with world-volume $AdS_2\times S^4$. The $S^4$ part
of the world-volume is inside $S^5$, so the blowing-up direction
($S^5$ in this case) is again common for both the chiral primary
and the Wilson loop. However, the droplet picture of an
anti-symmetric Wilson loop is a bit different from the giant
graviton: it corresponds to a line dividing the droplet. Still
there is a similarity between the droplet picture of giant
graviton and anti-symmetric Wilson loop, namely they are both
represented by objects inside the droplet. It would be nice to
push this analogy further.

We have noted that, when $\lambda$ is large, a symmetric
representation Wilson loop is practically identical to a large
winding number single trace Wilson loop.  This explains why they
can both be associated with the D3-brane.  We also observe that,
there is another phase which is stable for smaller $\lambda$ and a
phase transition at $\lambda\approx 5.5$.  In that phase, the free
energy resembles the D5-brane.  It is interesting that this case
has a D5-brane blowing up into the AdS$_5$ direction.

We also considered the OPE between Wilson loop and chiral primary.
At least technically, our normal matrix model gives an efficient way
to compute this OPE. We found a peculiar selection rule for the OPE.
For example, the Wilson loop in the fundamental representation
couples only to the chiral primary with single hook representation.
Similarly, the $K^{\rm th}$ anti-symmetric loop couples to the
chiral primary with at most $K$ hooks. We should emphasize that this
selection rule is exact. The meaning of selection rule in the bulk
$AdS$ side is not clear to us. It would be nice to understand this
from the string theory side.

Recently, the supergravity dual of the 1/2-BPS Wilson loop has been
constructed \cite{Yamaguchi:2006te,Lunin:2006xr}. The solution is
characterized by a boundary value problem in two-dimensions, and the
one-dimensional boundary line is divided into black and white
regions. It is suggested \cite{Yamaguchi:2006te} that these regions
corresponds to the eigenvalue distribution of a hermitian matrix
model. It would be nice to clarify this relation, if any. Our normal
matrix model description  seems to suggest the existence of M-theory
version of the supergravity dual, where the boundary region is
two-dimensional and the boundary data are droplets.

We mainly considered the Wilson loop in the representation whose
Young diagram has a number of boxes of order $N$. We expect that
when the number of boxes becomes of order $N^2$, the dual object
can no longer be regarded as a probe of the fixed background
$AdS_5\times S^5$, rather the geometry is significantly deformed.
In our droplet picture, the effect of such big Wilson loop will be
to deform the shape of the circular droplet. For example, the
anti-symmetric representation of the large winding loop $\bra
W_{A_K}^{(w)}\ket$ might correspond to two disconnected droplets
when $Kw\sim N^2$. It would be interesting to find the eigenvalue
distribution for this kind of large Wilson loop.

\end{document}